\documentclass[conference]{IEEEtran}
\IEEEoverridecommandlockouts

\usepackage{cite}
\usepackage{amsmath,amssymb,amsfonts}
\usepackage{algorithmic}
\usepackage{graphicx}
\usepackage{textcomp}
\usepackage{xcolor}
\usepackage{amsmath}
\usepackage{amsthm}
\usepackage{tikz}
\usepackage{subcaption}
\usetikzlibrary{arrows.meta, positioning}

\DeclareMathOperator{\diag}{\mathrm{diag}}

\DeclareMathOperator{\col}{\mathrm{col}}

\newtheorem{theorem}{Theorem}[]

\newtheorem{proposition}{Proposition}[]
\newtheorem{corollary}{Corollary}[theorem]
\newtheorem{lemma}{Lemma}[]
\newtheorem{definition}{Definition}[]
\newtheorem{assumption}{Assumption}
\newtheorem{remark}{Remark}

\newtheorem{problem}{Problem}

\def\BibTeX{{\rm B\kern-.05em{\sc i\kern-.025em b}\kern-.08em
    T\kern-.1667em\lower.7ex\hbox{E}\kern-.125emX}}
\begin{document}

\title{Scalable Practical Synchronization of Heterogeneous Linear Systems
}

\author{
\IEEEauthorblockN{
Alex Ardelean\IEEEauthorrefmark{1},
Giacomo Casadei\IEEEauthorrefmark{1},
Paolo Frasca\IEEEauthorrefmark{1}
}

\IEEEauthorblockA{\IEEEauthorrefmark{1}
Univ. Grenoble Alpes, CNRS, Inria, Grenoble INP, GIPSA-lab, 38000 Grenoble, France \\
alex.ardelean@grenoble-inp.fr
}
}

\maketitle

\begin{abstract}
    We study practical state synchronization of heterogeneous linear multi-agent systems over directed graphs using static diffusive state feedback. Persistent mismatches between the agent dynamics generally preclude exact synchronization and are treated as bounded disturbances within an Input-to-State Stability (ISS) framework. We derive a scalable synthesis procedure based on fixed-dimension linear matrix inequalities (LMIs), whose dimension is independent of the number of agents. Local robust dissipation conditions are combined through a small-gain argument to obtain global ISS guarantees and explicit bounds on the synchronization error. The framework is formulated in a unified manner for continuous- and discrete-time systems, while highlighting a fundamental difference in disturbance attenuation between the two domains. In particular, exact ZOH discretization of a continuous-time controller need not preserve closed-loop stability, motivating a native discrete-time synthesis. Numerical experiments on heterogeneous harmonic oscillators illustrate the proposed approach and the continuous/discrete-time distinction.
\end{abstract}


\section{Introduction}

Synchronization is a fundamental objective in the collective control of networked dynamical systems \cite{dorfler2014synchronization}, with widespread applications spanning distributed power grids \cite{blaabjerg2006overview,dorfler2014synchronization}, robotic swarms \cite{barcis2020sandsbots, schranz2020swarm}, and sensor networks \cite{sivrikaya2004time, ganeriwal2003timing}. 
The problem is to design distributed coupling laws that drive a group of agents toward a common behavior, by using only local information exchange.

While the synchronization of identical linear agents has been well understood for two decades~\cite{tuna2008synchronizing, scardovi2008synchronization, li2009consensus, hengster2013synchronization}, extending these results to heterogeneous dynamics is fundamentally more challenging. 
As dictated by the internal model principle \cite{francis1976internal, wieland2010static, wieland2011internal}, exact asymptotic synchronization necessitates that each agent embeds a common internal model of the synchronization trajectory. 
To satisfy this requirement, two main design paradigms are typically pursued: augmenting the agents with dynamic compensators that supply the missing model \cite{kim2010output, su2011cooperative, isidori2014robust}, including for CT and DT agents alike, $H_\infty$-based dynamic protocols embedding an internal model \cite{li2018cooperative}, or relying on static diffusive couplings when the internal model is already implicitly present in the agents' intrinsic dynamics \cite{seyboth2012static,seyboth2015robust}. 
When the internal model is unknown to all agents, it can alternatively be learned online via distributed adaptive schemes, at the cost of an adaptive compensator and of some a priori structural knowledge on the agents' class of dynamics \cite{baldi2019leaderless}.

Conversely, when the internal model structure is absent, perfect asymptotic synchronization is structurally unattainable \cite{wieland2010static, wieland2011internal,panteley2017synchronization,kim2015robustness,lunze2022synchronised}. 
In this case, the residual plant heterogeneity acts as a persistent disturbance on the network, preventing the synchronization error from converging to zero. Although dynamic controllers can recover exact synchronization, their increased order and complexity limit their applicability in resource-constrained settings.
This limitation motivates the study of \emph{practical synchronization}, where the synchronization error is not required to vanish but instead to remain ultimately bounded, with the steady-state bound scaling with the magnitude of the parametric mismatch. 
For diffusively coupled nonlinear agents satisfying a QUAD condition, this notion was shown to be achievable with an arbitrarily small ultimate bound by sufficiently increasing the algebraic connectivity of the coupling graph \cite{montenbruck2015practical}.
Here, we instead formalize practical synchronization within the Input-to-State Stability (ISS) framework \cite{dashkovskiy2006construction, dashkovskiy2010small} by treating the plant heterogeneity as an exogenous disturbance to the error dynamics, which allows us to derive explicit, LMI-based synthesis conditions for the coupling gain rather than relying solely on high-gain connectivity.
Beyond the single-system case, networks of interconnected ISS subsystems can in turn be certified as globally ISS through \emph{small-gain} arguments \cite{dashkovskiy2010small, geiselhart2014relaxed}, which is the route pursued in this paper to lift local, agent-level robustness guarantees to a global network-level statement.
A concurrent line of work \cite{hong2025small} has combined a small-gain argument with an LMI-based synthesis for the exact consensus without an internal model, relies on an edge-structured (rather than common) feedback gain, and requires solving a BMI via an iterative scheme whose size grows with the number of network edges. In contrast, our fixed-dimension LMIs depend only on the local vertex degree, preserving synthesis complexity independent of the network size.

Motivated by the challenges and tools highlighted in this literature, the present paper investigates the practical synchronization of heterogeneous linear multi-agent systems via static diffusive coupling. The main contributions are as follows.
\begin{itemize}
    \item We reformulate leader-following practical synchronization of heterogeneous linear systems under static diffusive state feedback as an ISS problem, explicitly characterizing the effect of persistent plant mismatch.
    \item We derive robust fixed-dimension LMIs for synthesis of a common feedback gain, whose dimension does not depend on the number of agents. The result relies on a common quadratic storage function and a robust treatment of both parametric uncertainty and network degree.
    \item We combine the local certificates with a max-type ISS Lyapunov function and a small-gain condition to obtain global synchronization guarantees and explicit error bounds for directed networks.
    \item We establish a unified CT/DT formulation and clarify the different disturbance-attenuation limits imposed by the two time domains. It turns out that,
    under suitable conditions on the system dynamics and controller structure, the ultimate error can be made arbitrarily small via high-gain design for continuous-time systems, whereas discrete-time systems face an inherent structural limit.
    \item We further show that ZOH discretization of a CT-designed controller need not preserve stability and motivate native DT synthesis.
\end{itemize}


Our approach significantly differs from previous work. Indeed,
for networks composed of identical (homogeneous) linear agents, a classical and elegant approach to analyze synchronization relies on the diagonalization of the network topology, typically represented by the Laplacian matrix \cite{olfati2007consensus, ren2008distributed, scardovi2008synchronization}. By projecting the collective dynamics onto the eigenspace of the graph, the network error equations can be successfully decoupled into $N$ independent, isolated subsystems. This paradigm effectively reduces the network-wide coordination problem to a set of standalone decentralized stability conditions. In sharp contrast, this decoupling mechanism breaks down in the heterogeneous setting investigated in this work. Even if a coordinate transformation is applied to the graph Laplacian, the parametric mismatches among the individual plant matrices act as persistent cross-coupling terms within the transformed state space \cite{trentelman2013robust,jongsma2016robust,li2018robust}. Consequently, the network error dynamics remain fundamentally interconnected, meaning that diagonalization fails to simplify or isolate the subsystems. This structural challenge formally motivates our departure from classical diagonalization techniques, justifying the adoption of robust control tools such as small-gain analysis and dissipative interconnected system theory \cite{dashkovskiy2010small, geiselhart2014relaxed} to establish practical synchronization.

The remainder of this paper is organized as follows. Section~\ref{section:preliminaries} recalls essential preliminary concepts regarding Input-to-State Stability (ISS) and small-gain theory. Section~\ref{section:problem_formulation} introduces the problem formulation and formalizes the heterogeneous multi-agent network. Section~\ref{sec:main_results} details the proposed solution, providing the control synthesis and global stability proofs. Section~\ref{section:numerical_example} validates the theoretical findings through a numerical example, and finally, Section~\ref{section:conclusion} draws the concluding remarks.

\subsection*{Notation}
Throughout this paper, we employ the following notation:

\begin{itemize}
    \item Sets and Scalars: $\mathbb{R}$ and $\mathbb{N}$ denote the sets of real and natural numbers, respectively. $\mathbb{R}_{\geq 0}$ denotes the set of non-negative real numbers. For a scalar $x \in \mathbb{R}$, $|x|$ denotes its absolute value. $\mathbb{T}$ denotes the unified time domain, where $\mathbb{T} = \mathbb{R}_{\geq 0}$ for continuous-time (CT) systems and $\mathbb{T} = \mathbb{N}$ for discrete-time (DT) systems.

    \item Unified Systems Framework: To describe CT and DT linear systems within a unified framework, we introduce the operator $\delta$. When applied to a state vector $x$, it is defined as:
    \begin{equation*}
    \delta x =
    \begin{cases} 
        \frac{dx(t)}{dt}, & \text{for CT systems,} \\
        x[k+1], & \text{for DT systems.}
    \end{cases}
    \end{equation*}
    The general LTI system is then expressed as $\delta x = Ax + Bu$. When applied to a scalar storage function $V$, $\delta V$ denotes its temporal variation, i.e., $\delta V = \dot{V}$ for CT and $\delta V = V[k+1] - V[k]$ for DT.

    \item Vectors and Matrices: $I_n$ represents the $n \times n$ identity matrix. The standard Euclidean norm is denoted by $\| \cdot \|$. For a set of vectors $x_1, \dots, x_N$, their vertical concatenation is $\operatorname{col}(x_1, \dots, x_N) = [x_1^\top \, \dots \, x_N^\top]^\top$. For any square matrix $A$, $\operatorname{spec}(A)$ and $\rho(A)$ denote its spectrum and spectral radius, respectively. The symmetric part is $\operatorname{sym}(A) = A^\top + A$. The Kronecker product is denoted by $\otimes$. The block diagonal matrix is written as $\operatorname{diag}(A_1, \dots, A_N)$.

    \item Matrix Inequalities: The sets of $n \times n$ symmetric positive definite and positive semi-definite real matrices are $\mathbb{S}^n_{\succ 0}$ and $\mathbb{S}^n_{\succeq 0}$. The notation $A \succ 0$ ($A \succeq 0$) indicates that the symmetric matrix $A$ is positive definite (semi-definite). For symmetric matrices $A$ and $B$, $A \succ B$ means $A - B \succ 0$. Element-wise strictly positive and non-negative matrices are denoted by $A > 0$ and $A \ge 0$, respectively.

    \item Comparison Functions: We adopt the standard comparison functions:
    \begin{itemize}
        \item A function $\alpha: \mathbb{R}_{\geq 0} \to \mathbb{R}_{\geq 0}$ is of class $\mathcal{K}$ if it is strictly increasing and $\alpha(0) = 0$.
        \item A function $\alpha \in \mathcal{K}$ is of class $\mathcal{K}_\infty$ if it is unbounded ($\alpha(s) \to \infty$ as $s \to \infty$).
        \item A function $\beta: \mathbb{R}_{\geq 0} \times \mathbb{T} \to \mathbb{R}_{\geq 0}$ is of class $\mathcal{KL}$ if $\beta(\cdot, t) \in \mathcal{K}$ for fixed $t$, and $\beta(s, \cdot)$ is decreasing and tends to $0$ as $t \to \infty$ for fixed $s > 0$.
    \end{itemize}
\end{itemize}

\section{Preliminaries on Input-to-State Stability and Small-Gain Theory for Linear Systems}

\label{section:preliminaries}


In this section, we recall the essential ISS and small-gain notions and the related tools that are needed in our work, referring to \cite{jiang1996lyapunov,dashkovskiy2006construction,dashkovskiy2007iss,dashkovskiy2010small} for the general theory of ISS for interconnected systems. As the systems considered throughout this paper are linear, we present here a linear specialization of this general theory, tailored to the needs of the subsequent sections.

Consider a linear system in the unified $\delta$-operator framework
\begin{equation}
    \label{eq:general_system}
    \delta q = Sq+Rp
\end{equation}
where $q\in\mathbb{R}^n$ is the state, $p\in\mathbb{R}^m$ is the input, and $S\in\mathbb{R}^{n\times n}$, $R\in\mathbb{R}^{n\times m}$. Let $q(t,q_0,p)$ denote the trajectory at time $t\in\mathbb{T}$ from $q_0$ under input $p$.

\begin{definition}[Input-to-State Stability]
    \label{def:iss}
    System \eqref{eq:general_system} is Input-to-State Stable (ISS) if there exist $\beta\in\mathcal{KL}$ and $\gamma\in\mathcal{K}$ such that, for any $q_0\in\mathbb{R}^n$ and any bounded input $p$,
    \begin{equation}
        \label{eq:iss_bound}
        \|q(t,q_0,p)\| \le \beta(\|q_0\|,t) + \gamma\left(\sup_{t\in\mathbb{T}}\|p(t)\|\right), \quad \forall t\in\mathbb{T}.
    \end{equation}
\end{definition}
The function $\gamma$ is the ISS gain of the system. For the linear system \eqref{eq:general_system}, ISS is equivalent to asymptotic stability of $S$ (in the sense dictated by the chosen $\delta$-operator), and this property is in turn equivalent to the existence of an ISS-Lyapunov function, i.e. a continuous $V$ satisfying suitable positivity bounds (see \cite[eq. 5]{jiang1996lyapunov}) and a dissipation inequality (see \cite[eq. 4-5]{jiang1996lyapunov}).

We now specialize these notions and facts to the interconnected, linear setting relevant to this paper. Consider $N$ subsystems in the unified $\delta$-operator framework
\begin{equation}
    \label{eq:subsystem_dynamics}
    \delta z_i = F_i z_i + \sum_{j \neq i} G_{ij} x_j + C_i u_i, \quad i = 1, \dots, N,
\end{equation}
with $z_i\in\mathbb{R}^{n_i}$, $z=\mathrm{col}(z_1,\dots,z_N)\in\mathbb{R}^n$, local input $u_i\in\mathbb{R}^{m_i}$, and matrices $F_i\in\mathbb{R}^{n_i\times n_i}$, $G_{ij}\in\mathbb{R}^{n_i\times n_j}$, $C_i\in\mathbb{R}^{n_i\times m_i}$.

\begin{definition}[ISS-Lyapunov function]
    \label{def:iss_lyap_subsystem}
    A continuous function $W_i:\mathbb{R}^{n_i}\to\mathbb{R}_{\ge 0}$ is an ISS-Lyapunov function for subsystem $i$ if there exist $\underline\alpha_i,\overline\alpha_i,\alpha_i\in\mathcal{K}_\infty$, class $\mathcal K$ functions $\chi_{ij}$ ($j\ne i$, with $\chi_{ii}\equiv0$), and $\chi_{iu}\in\mathcal K$ such that
    \begin{equation}
        \label{eq:subsystem_bounds}
        \underline\alpha_i(\|z_i\|)\le W_i(x_i) \le \overline\alpha_i(\|z_i\|),
    \end{equation}
    \begin{multline}
\label{eq:subsystem_implication}
         \|z_i\|\ge\max\Big\{\max_{j\ne i} \chi_{ij}(\|z_j\|),\ \chi_{iu}(\|u\|)\Big\} \implies\\ \delta W_i(z,u)\le-\alpha_i(\|z_i\|).
    \end{multline}
\end{definition}
An equivalent (see \cite[Remark 2.2]{jiang1996lyapunov}), often more practical, dissipative characterization replaces \eqref{eq:subsystem_implication} with
\begin{equation}
    \delta W_i(z,u)\le-\alpha_{D,i}(\|z_i\|)+\sum_{j\ne i} \gamma_{ij}(\|z_j\|)+\sigma_i(\|u\|)    \label{eq:subsystem_dissipativity}
\end{equation}
for some $\alpha_{D,i}\in\mathcal{K}_\infty$, class $\mathcal K$ functions $\gamma_{ij}$ ($\gamma_{ii}\equiv 0$), and $\sigma_i\in\mathcal{K}$. Indeed, fixing $\theta\in(0,1)$ and writing $\alpha_{D,i}(\|z_i\|)=(1-\theta)\alpha_{D,i}(\|z_i\|)+\theta\alpha_{D,i}(\|z_i\|)$, the second term can be split equally among the $N$ terms on the right-hand side of \eqref{eq:subsystem_dissipativity}: whenever $\|z_i\|\ge\max\{\max_{j\ne i}\chi_{ij}(\|z_j\|),\chi_{iu}(\|u\|)\}$ with $\chi_{ij}(s):=\alpha_{D,i}^{-1}(\tfrac{N}{\theta}\gamma_{ij}(s))$ and $\chi_{iu}(s):=\alpha_{D,i}^{-1}(\tfrac{N}{\theta}\sigma_i(s))$, each term $\tfrac{\theta}{N}\alpha_{D,i}(\|z_i\|)$ dominates the corresponding $\gamma_{ij}(\|z_j\|)$ (resp. $\sigma_i(\|u\|)$), so that \eqref{eq:subsystem_dissipativity} reduces to \eqref{eq:subsystem_implication} with $\alpha_i(s):=(1-\theta)\alpha_{D,i}(s)$.

In the linear-gain case relevant to this paper, $\gamma_{ij}(s) = \gamma_{ij}s$ for constants $\gamma_{ij}\ge 0$ ($\gamma_{ii}=0$), collected into the gain matrix
\begin{equation}
    \label{eq:gain_matrix}
    \Gamma = 
    \begin{bmatrix}
        0 & \gamma_{12} & \dots & \gamma_{1N} \\
        \gamma_{21} & 0 & \dots & \gamma_{2N} \\
        \vdots & \vdots & \ddots & \vdots \\
        \gamma_{N1} & \gamma_{N2} & \dots & 0
    \end{bmatrix} \in \mathbb{R}_{\ge0}^{N\times N}.    
\end{equation}
If every $W_i$ is an ISS-Lyapunov function for subsystem $i$ and
\begin{equation}
    \label{eq:spectral_radius}
    \rho(\Gamma) < 1
\end{equation}
then the interconnected system \eqref{eq:subsystem_dynamics} is ISS with respect to $u$ (see \cite[Corollary 7]{dashkovskiy2007iss}).
Moreover, under \eqref{eq:spectral_radius}, there exists $c=[c_1,\dots,c_N]^\top\in\mathbb{R}_{>0}^N$ with $\Gamma c<c$, and
\begin{equation}
    W(x)=\max_{i\in\{1,\dots,N\}}\left\{\frac{1}{c_i}W_i(z_i)\right\}
    \label{eq:global_lyap_max}
\end{equation}
is a global ISS-Lyapunov function for \eqref{eq:subsystem_dynamics} (see \cite[Lemma 5]{dashkovskiy2006construction}).

\begin{remark}[Construction of $c$]
\label{rem:construction_c}
If $\Gamma$ is irreducible (i.e., the interconnection graph is strongly connected), the Perron--Frobenius theorem applies directly and guarantees the existence of a strictly positive eigenvector $c\in\mathbb{R}_{>0}^N$ such that $\Gamma c = \rho(\Gamma) c < c$. When $\Gamma$ is reducible, a strictly positive $c$ can be built from the M-matrix property (see \cite[Lemma 2.0.1]{ruffer2007monotone}) of $I-\Gamma$: since $\rho(\Gamma)<1$, $(I-\Gamma)^{-1}=\sum_{k\ge0}\Gamma^k\ge0$, and $c=(I-\Gamma)^{-1}\mathbf{1}_N$ satisfies $c\in\mathbb{R}_{>0}^N$, $\Gamma c<c$. An alternative perturbation-based construction is discussed in \cite{dashkovskiy2010small} by using \cite[Section 15.5, Th. 1]{gantmakher2000theory}; we employ this construction in Remark \ref{rem:optimal_c} to handle the reducible case when optimizing the choice of $c$.
\end{remark}

\section{Problem Formulation}

\label{section:problem_formulation}

In this section, the synchronization problem for heterogeneous linear multi-agent systems evolving over a unified time domain $\mathbb{T}$ is formally introduced. The problem is cast into a form suitable for analysis via Input-to-State Stability (ISS) framework and the small-gain theorem. In particular, the synchronization objective is expressed in terms of a network of interconnected error dynamics, allowing the direct deployment of localized ISS-Lyapunov conditions to establish sufficient conditions for practical synchronization.

\subsection{Network Model}

Consider a multi-agent system consisting of one virtual leader, indexed by $0$, and $N$ followers, indexed from $1$ to $N$. The dynamics of all agents are described within the unified $\delta$-operator framework over the time domain $\mathbb{T}$. The autonomous virtual leader evolves according to:
\begin{equation}
    \delta x_{0} = A_{0}x_{0},
    \label{eq:leader_dynamics}
\end{equation}
where $x_{0} \in \mathbb{R}^{n}$ denotes the leader state and $A_{0} \in \mathbb{R}^{n \times n}$ is the system matrix.

Each follower agent $i \in \{1, \dots, N\}$ is characterized by heterogeneous linear dynamics of the form:
\begin{equation}
    \delta x_{i} = A_{i}x_{i} + B u_{i},
    \label{eq:follower_dynamics}
\end{equation}
where $x_{i} \in \mathbb{R}^{n}$ is the state of agent $i$, $u_{i} \in \mathbb{R}^{m}$ is the control input, $B \in \mathbb{R}^{n \times m}$ is a common input matrix, and $A_{i} \in \mathbb{R}^{n \times n}$ are the individual system matrices.

To formalize the plant parameter mismatches, we introduce a nominal matrix $A \in \mathbb{R}^{n \times n}$ utilized for the control design, and decompose the agent matrices as:
\begin{equation}
    A_{i} = A + \Delta_{i}, \quad \forall i \in \{0, 1, \dots, N\},
    \label{eq:matrix_decomposition}
\end{equation}
where $\Delta_{i} \in \mathbb{R}^{n \times n}$ represents the modeling heterogeneity of the $i$-th agent.

\begin{assumption}[Bounded Heterogeneity]
\label{ass:bounded_heterogeneity}
There exists a known constant $\bar{\delta} \ge 0$ such that the induced matrix norm of the heterogeneities satisfies:
\begin{equation}
    \|\Delta_{i}\| \le \bar{\delta}, \quad \forall i \in \{0, 1, \dots, N\} \text{}.
\end{equation}
\end{assumption}

The information exchange topology among the $N+1$ agents is modeled by a directed graph $\mathcal{G} = (\mathcal{V}, \mathcal{E})$ without self-loops. Here, $\mathcal{V} = \{0, 1, \dots, N\}$ represents the node set and $\mathcal{E} \subseteq \mathcal{V} \times \mathcal{V}$ is the edge set such that $(i,i) \notin \mathcal{E}$ for all $i \in \mathcal{V}$. A directed edge $(i,j) \in \mathcal{E}$ indicates that agent $i$ can access the state information of agent $j$. The neighborhood of the $i$-th agent is defined as $\mathcal{N}_{i} = \{j \in \mathcal{V} : (i,j) \in \mathcal{E}\}$.

\begin{assumption}[Connectivity]
\label{ass:connectivity}
The directed communication graph $\mathcal{G}$ contains a directed spanning tree rooted at the leader node $0$.
\end{assumption}
This assumption structurally guarantees that the information generated by the virtual leader can propagate through the network to every follower. Under this topology condition, cooperative control objectives can be achieved, as established for instance in \cite[Theorem 1]{zoboli2026incremental}.

\begin{assumption}[Bounded and Non-Vanishing Leader Trajectory]
\label{ass:leader_trajectory}
The trajectory of the virtual leader $x_{0}(\tau)$ generated by \eqref{eq:leader_dynamics} is bounded and non-vanishing. That is, there exist constants $M \ge m > 0$ such that:
\begin{equation}
    \sup_{\tau \in \mathbb{T}} \|x_{0}(\tau)\| \le M \quad \text{and} \quad \limsup_{\tau \to \infty} \|x_{0}(\tau)\| \ge m.
\end{equation}
\end{assumption}
Assumption \ref{ass:leader_trajectory} ensures a mathematically and practically meaningful synchronization problem. Requiring a bounded reference trajectory prevents the followers from having to track unbounded signals, which is impractical under real-world constrained control inputs. Concurrently, the non-vanishing condition prevents the leader's state from asymptotically decaying to the origin. If $x_{0}(\tau)$ were to approach zero asymptotically, the practical synchronization problem would degenerate into standard asymptotic stabilization at the origin.

Let $\mathcal{A} = [a_{ij}] \in \mathbb{R}_{\ge 0}^{(N+1) \times (N+1)}$ (with $a_{ii}=0$) denote the adjacency matrix associated with $\mathcal{G}$. The network Laplacian matrix $L \in \mathbb{R}^{(N+1) \times (N+1)}$ is defined as:
\begin{equation}
    L := D - \mathcal{A},
    \label{eq:laplacian}
\end{equation}
where $D := \text{diag}(d_{0}, \dots, d_{N})$ is the in-degree diagonal matrix with entries $d_{i} := \sum_{j=0}^{N} a_{ij}$.

Each follower agent implements a localized diffusive control law based on relative state measurements:
\begin{equation}
    u_{i} = \sum_{j \in \mathcal{N}_{i}} a_{ij} K (x_{i} - x_{j}) = K \sum_{j=0}^{N} l_{ij} x_{j}, \quad i = 1, \dots, N,
    \label{eq:control_law}
\end{equation}
where $K \in \mathbb{R}^{m \times n}$ is a common feedback gain matrix to be designed, and $l_{ij}$ denotes the $(i,j)$-th entry of $L$.

By injecting the control policy \eqref{eq:control_law} into \eqref{eq:follower_dynamics} and defining the collective state vector $x = \text{col}(x_{0}, x_{1}, \dots, x_{N}) \in \mathbb{R}^{(N+1)n}$, the full closed-loop network system is written compactly as:
\begin{equation}
    \delta x = \left( I_{N+1} \otimes A + \Delta_{\text{tot}} + (L \otimes BK) \right) x,
    \label{eq:closed_loop_compact}
\end{equation}
where $\Delta_{\text{tot}} = \text{diag}(\Delta_{0}, \Delta_{1}, \dots, \Delta_{N}) \in \mathbb{R}^{(N+1)n \times (N+1)n}$.

\subsection{Error Dynamics}

To evaluate the cooperation and tracking performance across the multi-agent network, we define the local synchronization error of the $i$-th follower with respect to the virtual leader as:
\begin{equation}
    e_{i} := x_{i} - x_{0}, \quad i = 1, \dots, N.
    \label{eq:local_error_def}
\end{equation}

Taking the $\delta$-derivative along the trajectories of the follower dynamics \eqref{eq:follower_dynamics} and the leader dynamics \eqref{eq:leader_dynamics} yields:
\begin{equation}
    \delta e_{i} = A_{i}x_{i} + BK \sum_{j=0}^{N} a_{ij}(x_{i} - x_{j}) - A_{0}x_{0}.
    \label{eq:delta_error_initial}
\end{equation}
Substituting the plant parameter expansions $A_{i} = A + \Delta_{i}$ and $A_{0} = A + \Delta_{0}$ into \eqref{eq:delta_error_initial} results in:
\begin{equation}
    \delta e_{i} = A(x_{i} - x_{0}) + \Delta_{i}x_{i} - \Delta_{0}x_{0} + BK \sum_{j=0}^{N} a_{ij}(x_{i} - x_{j}).
    \label{eq:delta_error_expanded}
\end{equation}
By expressing the follower states through the synchronization error as $x_{i} = x_{0} + e_{i}$, the heterogeneity terms can be algebraically manipulated as follows:
\begin{equation}
    \Delta_{i}x_{i} - \Delta_{0}x_{0} = \Delta_{i}(x_{0} + e_{i}) - \Delta_{0}x_{0} = \Delta_{i}e_{i} + (\Delta_{i} - \Delta_{0})x_{0}.
    \label{eq:heterogeneity_manipulation}
\end{equation}
Letting $H_{i} := \Delta_{i} - \Delta_{0} \in \mathbb{R}^{n \times n}$ quantify the relative plant parameter mismatch between the $i$-th follower and the leader, the error equation simplifies to:
\begin{equation}
    \delta e_{i} = (A + \Delta_{i})e_{i} + BK \sum_{j=0}^{N} a_{ij}(x_{i} - x_{j}) + H_{i}x_{0}.
    \label{eq:delta_error_with_H}
\end{equation}

Noting that the relative state difference can be written entirely in terms of the local errors, i.e., $x_{i} - x_{j} = (e_{i} + x_{0}) - (e_{j} + x_{0}) = e_{i} - e_{j}$, the localized diffusive coupling term can be split into its individual and external components:
\begin{equation}
    BK \sum_{j=0}^{N} a_{ij}(x_{i} - x_{j}) = d_{i}BKe_{i} - BK \sum_{j=1}^{N} a_{ij}e_{j}.
    \label{eq:coupling_rewritten}
\end{equation}
Consequently, the local error dynamics for the $i$-th follower subsystem reduces to:
\begin{equation}
    \delta e_{i} = (A + \Delta_{i} + d_{i}BK)e_{i} - BK \sum_{j=1}^{N} a_{ij}e_{j} + H_{i}x_{0}.
    \label{eq:local_error_dynamics_final}
\end{equation}
The synchronization errors $e_{j}$ ($j \neq i$) act as the internal coupling variables, whereas the autonomous leader state $x_{0}$ behaves as the exogenous input driving the error growth.

By stacking the local errors into the aggregate network error vector $e = \text{col}(e_{1}, \dots, e_{N}) \in \mathbb{R}^{Nn}$, the global error system can be modeled compactly as:
\begin{equation}
    \delta e = \left[ (I_{N} \otimes A) + \Delta + (L_{22} \otimes BK) \right] e + Hx_{0},
    \label{eq:global_error_system}
\end{equation}
where $\Delta := \text{diag}(\Delta_{1}, \dots, \Delta_{N}) \in \mathbb{R}^{Nn \times Nn}$, $H := \text{col}(H_{1}, \dots, H_{N}) \in \mathbb{R}^{Nn \times n}$, and $L_{22} \in \mathbb{R}^{N \times N}$ represents the submatrix of the full network Laplacian $L$ corresponding exclusively to the interactions among the followers, per the standard graph partitioning:
\begin{equation}
    L = \begin{pmatrix} L_{11} & L_{12} \\ L_{21} & L_{22} \end{pmatrix},
    \label{eq:laplacian_partition}
\end{equation}
where the first row and column separate the leader node $0$ from the follower set.

\subsection{Practical Synchronization Objective}
\label{subsec:practical_synchronization}

Owing to the structure of the collective error dynamics \eqref{eq:global_error_system}, the cooperative tracking task can be framed within the Input-to-State Stability (ISS) paradigm, treating the leader state $x_0$ as the external disturbance input.

\begin{definition}[Practical Synchronization]
\label{def:practical_sync}
The multi-agent network is said to achieve practical synchronization if the aggregate error system \eqref{eq:global_error_system} is Input-to-State Stable (ISS) with respect to the input $x_0$. That is, there exist a class $\mathcal{KL}$ function~$\beta$ and a class $\mathcal{K}$ function~$\gamma$ such that, for any initial error configuration $e(\tau_0) \in \mathbb{R}^{Nn}$ and any bounded leader trajectory $x_0$, the trajectory solution satisfies:
\begin{equation}
    \|e(\tau)\| \le \beta(\|e(\tau_0)\|, \tau - \tau_0) + \gamma\left(\sup_{\tau_0 \le s \le \tau} \|x_0(s)\|\right),
    \label{eq:iss_definition}
\end{equation}
for all $\tau \ge \tau_0$ with $\tau, \tau_0 \in \mathbb{T}$.
\end{definition}

This formulation clarifies that under bounded leader references, the tracking error remains bounded and scales dynamically with the magnitude of $\|x_0\|$. Due to the persistent plant heterogeneity ($\Delta_i \neq \Delta_0$), perfect asymptotic synchronization ($\|e(\tau)\| \to 0$) is generally unattainable, motivating the practical synchronization goal.

\begin{problem}[Heterogeneous Linear Practical Synchronization]
\label{prob:main_problem}
Consider the multi-agent network defined by the leader node \eqref{eq:leader_dynamics} and the $N$ followers \eqref{eq:follower_dynamics}, under the bounded heterogeneity condition (Assumption \ref{ass:bounded_heterogeneity}), the communication graph topology configuration (Assumption \ref{ass:connectivity}), and the bounded and non-vanishing leader trajectory requirement (Assumption \ref{ass:leader_trajectory}). The objective is to synthesize a common feedback gain matrix $K \in \mathbb{R}^{m \times n}$ for the localized diffusive control protocol \eqref{eq:control_law} such that the resulting interconnected collective error dynamics \eqref{eq:global_error_system} is ISS with respect to the leader state $x_0$.
\end{problem}

\section{Main Results}

\label{sec:main_results}

In this section, we derive constructive conditions to synthesize the local feedback gain $K$ for the practical synchronization of heterogeneous multi-agent systems, as formulated in Problem \ref{prob:main_problem}. We employ the unified $\delta$-operator framework to provide a design approach valid for both continuous-time and discrete-time systems. We proceed in three steps. First, we provide general matrix inequality conditions guaranteeing dissipativity for a generic interconnection of linear subsystems (Lemma~\ref{lem:UnifiedDissipativityCondition}), and then adapt them to our setting through a scale-independent robust version tailored to the heterogeneous network structure (Lemma~\ref{lem:RobustUnifiedDissipativity}), which is subsequently linearized into tractable synthesis LMIs for the CT and DT domains (Propositions~\ref{prop:CT_LMI}--\ref{prop:DT_LMI}). Finally, we combine these local dissipativity certificates with the small-gain theorem to establish the main theorem, providing a solution to Problem~\ref{prob:main_problem} together with explicit ISS bounds on the synchronization error (Theorem~\ref{th:main_theorem}).

\subsection{Dissipativity Conditions}
\label{subsec:dissipativity}

We characterize the local stability properties of each follower agent by considering the general interconnected linear dynamics recalled in \eqref{eq:subsystem_dynamics}. By defining the aggregated vector of neighboring states $z_{-i} := \operatorname{col}(\{z_j\}_{j \neq i}) \in \mathbb{R}^{(N-1)n}$ and the stacked coupling matrix $\mathcal{G}_i := [G_{i1}, \dots, G_{i,i-1}, G_{i,i+1}, \dots, G_{iN}] \in \mathbb{R}^{n \times (N-1)n}$, the isolated subsystem dynamics can then be expressed in the standard interconnected form:
\begin{equation}
    \label{eq:subsystem_compact_generic}
    \delta z_i = F_i z_i + \mathcal{G}_i z_{-i} + C_i u_i.
\end{equation}
Furthermore, we define the block-diagonal interconnection gain matrix:
\[
    \Gamma_i := \operatorname{diag} \left( \gamma_{i1} P_1, \dots, \gamma_{i,i-1} P_{i-1}, \gamma_{i,i+1} P_{i+1}, \dots, \gamma_{iN} P_N \right),
\]
and introduce the structural matrix $\Phi$, defined to unify continuous-time (CT) and discrete-time (DT) domains:
\begin{equation}
    \label{eq:phi_definition}
    \Phi =
    \begin{cases}
        \begin{bmatrix} 0 & 1 \\ 1 & 0 \end{bmatrix}, & \text{continuous-time (CT)}, \\[12pt]
        \begin{bmatrix} -1 & 0 \\ 0 & 1 \end{bmatrix}, & \text{discrete-time (DT)}.
    \end{cases}
\end{equation}

\begin{lemma}[Unified Dissipativity Condition]
\label{lem:UnifiedDissipativityCondition}
Consider subsystem $i$ described by \eqref{eq:subsystem_compact_generic}. For given scalars $\alpha_i > 0$, $\eta_i > 0$, $\gamma_{ij} \ge 0$, and the storage function $V_i(z_i) = z_i^\top P_i z_i$ with $P_i = P_i^\top \succ 0$, the dissipation inequality
\begin{equation}
    \label{eq:scalar_dissipation}
    \delta V_i \le -\alpha_i V_i(z_i) + \sum_{j \neq i} \gamma_{ij} V_j(z_j) + \eta_i |u_i|^2
\end{equation}
holds if and only if the following matrix inequality is satisfied:
\begin{equation}
    \label{eq:MI_dissipation}
    \begin{bmatrix} I & A_i^\top \\ 0 & \mathcal{B}_i^\top \\ 0 & E_i^\top \end{bmatrix} (\Phi \otimes P_i) \begin{bmatrix} I & 0 & 0 \\ A_i & \mathcal{B}_i & E_i \end{bmatrix} + \begin{bmatrix} \alpha_i P_i & 0 & 0 \\ 0 & -\Gamma_i & 0 \\ 0 & 0 & -\eta_i I \end{bmatrix} \preceq 0.
\end{equation}
\end{lemma}

\begin{proof}
For a quadratic function $V_i(z_i) = z_i^\top P_i z_i$, a direct computation of its variation yields the unified quadratic representation:
\begin{equation}
    \label{eq:deltaV_unified_quad}
    \delta V_i = \begin{pmatrix} z_i \\ \delta z_i \end{pmatrix}^{\!\top} (\Phi \otimes P_i) \begin{pmatrix} z_i \\ \delta z_i \end{pmatrix},
\end{equation}
where $\Phi$ is defined as in \eqref{eq:phi_definition}. 

Using the subsystem dynamics \eqref{eq:subsystem_compact_generic}, define the stacked vector $\xi_i := \operatorname{col}(z_i, z_{-i}, u_i)$ and the lifting matrix $\mathcal{M}_i := \begin{bmatrix} I & 0 & 0 \\ F_i & \mathcal{G}_i & C_i \end{bmatrix}$. Then $\operatorname{col}(z_i, \delta z_i) = \mathcal{M}_i \xi_i$, which implies $\delta V_i = \xi_i^\top \mathcal{M}_i^\top (\Phi \otimes P_i) \mathcal{M}_i \xi_i$. 

The inequality \eqref{eq:scalar_dissipation} can be rewritten as:
\[
    \delta V_i + \alpha_i V_i(z_i) - \sum_{j \neq i} \gamma_{ij} V_j(z_j) - \eta_i u_i^\top u_i \le 0.
\]
Substituting the quadratic forms $V_i(z_i) = \xi_i^\top \operatorname{diag}(P_i, 0, 0) \xi_i$, $\sum \gamma_{ij} V_j = \xi_i^\top \operatorname{diag}(0, \Gamma_i, 0) \xi_i$, and $u_i^\top u_i = \xi_i^\top \operatorname{diag}(0, 0, I) \xi_i$, we obtain:
\[
    \xi_i^\top \left( \mathcal{M}_i^\top (\Phi \otimes P_i) \mathcal{M}_i + \operatorname{diag}(\alpha_i P_i, -\Gamma_i, -\eta_i I) \right) \xi_i \le 0.
\]
Since this quadratic inequality must hold for all arbitrary vectors $\xi_i \in \mathbb{R}^{n + (N-1)n + m}$, it is satisfied if and only if the corresponding symmetric matrix is negative semi-definite, which is exactly the matrix inequality presented in \eqref{eq:MI_dissipation}.
\end{proof}

To decouple the dissipativity conditions from the network size $N$, we assume an interaction structure where the interconnection matrices decompose as $G_{ij} = a_{ij}B_i$, with $a_{ij} \ge 0$ being the scalar coupling weight from subsystem $j$ to subsystem $i$, and $B_i$ the input matrix of subsystem~$i$. We introduce the local aggregated interconnection variable:
\begin{equation}
    v_i := \sum_{j \neq i} a_{ij} x_j,
\end{equation}
which captures the total network influence on subsystem $i$ while preserving the state dimension. We restrict attention to the case where the dynamics of each subsystem are subject to a structured parameter variation $\beta_i T$, with scalar $\beta_i$ confined to the interval $[\beta_{\min}, \beta_{\max}]$ and known shape matrix $T \in \mathbb{R}^{n \times n}$, and to an unstructured heterogeneity $\Delta_i \in \mathbb{R}^{n \times n}$, described by the compact form
\begin{equation}
 \label{eq:robust_subsystem_dynamics}
    \delta x_i = (A + \beta_i T + \Delta_i) x_i + B_i v_i + E_i u_i,
\end{equation}
a structure chosen so as to later match, and thus be directly applicable to, the local error dynamics \eqref{eq:local_error_dynamics_final} of our synchronization problem. Here $\Delta_i$ represents an unstructured, norm-bounded heterogeneous uncertainty such that $\|\Delta_i\| \le \bar{\delta}$. To ensure the resulting matrix inequality is completely independent of the network size, we assume a homogeneous Lyapunov matrix $P = P^\top \succ 0$ across all subsystems: should $P_i$ be  allowed to vary with $i$, the interconnection block $\Gamma_i$ would in general require one copy of $P_j$ per neighbor $j$ of subsystem $i$, so that the size of the resulting matrix inequality would grow with the local degree of the network: precisely the scaling with $N$ we want to avoid.

\begin{lemma}[Robust scale-independent dissipativity condition]
\label{lem:RobustUnifiedDissipativity}
Consider subsystem $i$ described by \eqref{eq:robust_subsystem_dynamics} with parameter $\beta_i \in [\beta_{\min}, \beta_{\max}]$ and uncertainty $\|\Delta_i\| \le \bar{\delta}$. For given design scalars $\alpha > 0$, $\gamma > 0$, and $\eta > 0$, the dissipation inequality
\begin{equation}
    \label{eq:size_indep_scalar_dissipation_robust}
    \delta V_i(x_i) \le -\alpha V_i(x_i) + \gamma d_i \sum_{j \neq i} a_{ij} V_j + \eta \|u_i\|^2
\end{equation}
holds for all $\beta_i \in [\beta_{\min}, \beta_{\max}]$ and all $\|\Delta_i\| \le \bar{\delta}$ if there exists a scalar $\mu > 0$ such that the following fixed-dimension matrix inequality
\begin{equation}
    \label{eq:unified_robust_LMI}
    \resizebox{\columnwidth}{!}{$
    \begin{bmatrix} I & (A+\beta_i T)^\top \\ 0 & B_i^\top \\ 0 & E_i^\top \\ 0 & I \end{bmatrix} (\Phi \otimes P) \begin{bmatrix} I & 0 & 0 & 0 \\ A+\beta_i T & B_i & E_i & I \end{bmatrix} + \begin{bmatrix} \alpha P + \mu \bar{\delta}^2 I & 0 & 0 & 0 \\ 0 & -\gamma P & 0 & 0 \\ 0 & 0 & -\eta I & 0 \\ 0 & 0 & 0 & -\mu I \end{bmatrix} \preceq 0
    $}
\end{equation}
is satisfied at the vertices $\beta_i \in \{\beta_{\min}, \beta_{\max}\}$.
\end{lemma}

\begin{proof}
    Define the uncertain term as an additional exogenous input $w_i := \Delta_i x_i$. The bound $\|\Delta_i\|\le\overline\delta$ then gives the quadratic constraint:
    \[
        |w_i|^2 = x_i^\top \Delta_i^\top \Delta_i x_i \le \bar{\delta}^2 x_i^\top x_i \implies \bar{\delta}^2 x_i^\top x_i - w_i^\top w_i \ge 0,
    \]
    and the dynamics become exactly
    \[
        \delta x_i = (A + \beta_i T)x_i + B_i v_i + E_i u_i + I w_i.
    \]

    Collecting the augmented vector $\tilde{\xi}_i := \operatorname{col}(x_i, v_i, u_i, w_i)$, this reads $\col(x_i,\delta x_i)=\mathcal{N}_i\tilde{\xi}_i$ with $\mathcal{N}_i:=\begin{bmatrix} I & 0 & 0 & 0 \\ A+\beta_i T & B_i & E_i & I \end{bmatrix}$, so that $\delta V(x_i) = \tilde{\xi}_i^\top \mathcal{N}_i^\top (\Phi \otimes P) \mathcal{N}_i \tilde{\xi}_i$.

    For the interconnection term, since $P\succ 0$ admits $P^{\frac{1}{2}}$, writing $v_i^\top P v_i =\left\|\sum_{j\neq i}\sqrt{a_{ij}}\left(\sqrt{a_{ij}}P^{\frac{1}{2}}x_j\right)\right\|^2$ and applying Cauchy-Schwartz (using $a_{ij}\ge 0$) gives
    \[
        \begin{aligned}
            v_i^\top P v_i &\le \left( \sum_{j \neq i} \left(\sqrt{a_{ij}}\right)^2 \right) \left( \sum_{j \neq i} \left\| \sqrt{a_{ij}} P^{1/2} x_j \right\|^2 \right) \\
            &= \left( \sum_{j \neq i} a_{ij} \right) \left( \sum_{j \neq i} a_{ij} x_j^\top P x_j \right) = d_i \sum_{j \neq i} a_{ij} V_j.
        \end{aligned}
    \]

    Hence \eqref{eq:size_indep_scalar_dissipation_robust} is implied by the stricter condition 
    \[
        \delta V(x_i) + \alpha x_i^\top P x_i - \gamma v_i^\top P v_i - \eta \|u_i\|^2 \le 0.
    \]
    Expressed in matrix form with respect to the augmented vector $\tilde{\xi}_i$, this requirement becomes:
    \[
        \tilde{\xi}_i^\top \left[ \mathcal{N}_i^\top (\Phi \otimes P) \mathcal{N}_i + \operatorname{diag}(\alpha P, -\gamma P, -\eta I, 0) \right] \tilde{\xi}_i \le 0.
    \]
    for all $\tilde{\xi}_i$ satisfying $\tilde{\xi}_i^\top \operatorname{diag}(\bar{\delta}^2 I, 0, 0, -I) \tilde{\xi}_i \ge 0$. 

    By the S-Procedure (\cite{polik2007survey} Theorem 2.2) , this holds iff there exists $\mu>0$ such that
    \begin{multline*}
        \mathcal{N}_i^\top (\Phi \otimes P) \mathcal{N}_i + \operatorname{diag}(\alpha P, -\gamma P, -\eta I, 0) + \\ + \mu \operatorname{diag}(\bar{\delta}^2 I, 0, 0, -I) \preceq 0,
    \end{multline*}

    It remains to show that checking \eqref{eq:unified_robust_LMI} at $\beta_i\in\{\beta_{\min},\beta_{\max}\}$ suffices for all $\beta_i\in[\beta_{\min},\beta_{\max}]$. Write $\mathcal{N}_i(\beta_i) = \mathcal{N}_0 + \beta_i\mathcal{N}_T$ with $\mathcal{N}_T := \begin{bmatrix}0&0&0&0\\T&0&0&0\end{bmatrix}$; then $\mathcal{L}(\beta_i)$, the left-hand side of \eqref{eq:unified_robust_LMI}, is quadratic in $\beta_i$ with leading coefficient $\mathcal{Q} := \mathcal{N}_T^\top(\Phi\otimes P)\mathcal{N}_T$, all other terms being affine. In continuous time $\Phi\otimes P$ is off-diagonal, so $\mathcal{Q}=0$ and $\mathcal{L}(\beta_i)$ is affine in $\beta_i$; in discrete time $\Phi\otimes P = \operatorname{diag}(-P,P)$, so $\mathcal{Q} = \operatorname{diag}(T^\top PT,0,0,0) \succeq 0$, making $\beta_i\mapsto \tilde\xi^\top\mathcal{L}(\beta_i)\tilde\xi$ convex for every fixed $\tilde\xi$. In either case $\mathcal{L}(\beta_i)$ is convex on $[\beta_{\min},\beta_{\max}]$, so its maximum over this interval is attained at the endpoints; thus $\mathcal{L}(\beta_{\min})\preceq0$ and $\mathcal{L}(\beta_{\max})\preceq0$ are necessary and sufficient for $\mathcal{L}(\beta_i)\preceq0$ throughout the interval.

\end{proof}

\subsection{LMI Characterizations for Control Synthesis}

Starting from the local error dynamics \eqref{eq:local_error_dynamics_final}, and relabeling the leader-induced term $H_i x_0$ as a generic bounded disturbance $w_i$, the error dynamics of subsystem $i$ take the compact form
\begin{equation} \label{eq:error_dynamics}
    \delta e_i = (A + \Delta_i + d_i BK)e_i - BK \sum_{j \neq i}^N a_{ij} e_j + w_i.
\end{equation}
This expression instantiates the robust subsystem structure \eqref{eq:robust_subsystem_dynamics} of Lemma~\ref{lem:RobustUnifiedDissipativity} under the identification $\beta_i \equiv d_i$, $T \equiv BK$, $B_i \equiv -BK$, $E_i \equiv I$, $v_i \equiv \sum_{j\neq i} a_{ij}e_j$, and $u_i \equiv w_i$. The objective of this subsection is to provide tractable LMIs for the synthesis of the control gain $K$, obtained by specializing the robust scale-independent matrix inequality \eqref{eq:unified_robust_LMI} to this identification, for both the continuous-time and discrete-time domains.

\begin{proposition}[Continuous-Time LMI for dissipativity]
\label{prop:CT_LMI}
Consider the continuous-time error dynamics \eqref{eq:error_dynamics}. The system satisfies the robust scale-independent dissipation inequality \eqref{eq:size_indep_scalar_dissipation_robust} if there exist $X = X^\top \succ 0$, $Y$, and $\tau > 0$ such that, for all vertices $d_i \in \{d_{\min}, d_{\max}\}$, the following LMI holds:
\begin{equation}
    \label{eq:CT_LMI_final}
    \resizebox{\columnwidth}{!}{$
    \begin{bmatrix}
        \mathrm{sym}(AX + d_i BY) + \alpha X & -BY & I & \tau I & \bar{\delta} X \\
        -Y^\top B^\top & -\gamma X & 0 & 0 & 0 \\
        I & 0 & -\eta I & 0 & 0 \\
        \tau I & 0 & 0 & -\tau I & 0 \\
        \bar{\delta} X & 0 & 0 & 0 & -\tau I
    \end{bmatrix} \preceq 0,
    $}
\end{equation}
where the stabilizing feedback control gain is recovered via $K = Y X^{-1}$.
\end{proposition}

\begin{proof}
    Under the identification in \eqref{eq:error_dynamics}, the robust condition \eqref{eq:unified_robust_LMI} instantiates to a BMI in $(P,K)$, bilinear through the terms $PBK$. Applying the congruence transformation $\diag(P^{-1},P^{-1},I,I)$ and setting $X:=P^{-1}$, $Y=KX$ removes this coupling everywhere except for a residual quadratic uncertainty term $\mu\overline\delta^2 X^2$ in the $(1,1)$ block. Writing this term as $(\overline\delta X)(\mu^{-1}I)^{-1}(\overline\delta X)$ and applying the Schur complement appends a fifth block row and column, with $\overline\delta X$ at position $(1,5)/(5,1)$ and $\mu^{-1}I$ at $(5,5)$, leaving the $(1,1)$ block free of $\mu$. Finally, applying the congruence transformation $\diag(I,I,I,\tau I,I)$, maps $(1,4)/(4,1)$ blocks from $I$ to $\tau I$ and the $(4,4)$ block from $-\mu I$ to $-\mu\tau^2I=-\tau I$, while $(5,5)$ block is $-\mu^{-1}I=-\tau I$. 
\end{proof}

\begin{proposition}[Discrete-Time LMI for dissipativity]
\label{prop:DT_LMI}
Consider the discrete-time error dynamics \eqref{eq:error_dynamics}. The system satisfies the robust scale-independent dissipation inequality \eqref{eq:size_indep_scalar_dissipation_robust} if there exist $X = X^\top \succ 0$, $Y$, and $\tau > 0$ such that, for all vertices $d_i \in \{d_{\min}, d_{\max}\}$, the following LMI holds:
\begin{equation}
    \label{eq:DT_LMI_final}
    \resizebox{\columnwidth}{!}{$
    \begin{bmatrix}
        (\alpha - 1)X & 0 & 0 & X A^\top + d_i Y^\top B^\top & 0 & \bar{\delta} X \\
        0 & -\gamma X & 0 & -Y^\top B^\top & 0 & 0 \\
        0 & 0 & -\eta I & I & 0 & 0 \\
        A X + d_i B Y & -B Y & I & -X & \tau I & 0 \\
        0 & 0 & 0 & \tau I & -\tau I & 0 \\
        \bar{\delta} X & 0 & 0 & 0 & 0 & -\tau I
    \end{bmatrix} \preceq 0,
    $}
\end{equation}
where the stabilizing feedback control gain is recovered via $K = Y X^{-1}$.
\end{proposition}

\begin{proof}
    Under the identification in \eqref{eq:error_dynamics}, the discrete-time instantiation of \eqref{eq:unified_robust_LMI} contains the nonlinear term $(A+d_iBK)^\top P(A+d_iBK)$ in the $(1,1)$ block. Writing this as $(A+d_iBK)^\top PP^{-1}P(A+d_iBK)$ and applying the Schur complement introduces a fourth block row and column, with $P(A+d_iBK)$ at $(4,1)/(1,4)$ and $-P$ at $(4,4)$, reducing the $(1,1)$ block to $(\alpha-1)P+\mu\overline\delta^2I$. This yields a BMI in $(P,K)$ with five block rows and columns, bilinear through $P(A+d_iBK)$ and $PBK$.
    Applying the congruence transformation $\diag(P^{-1},P^{-1},I,P^{-1},I)$ and setting $X:=P^{-1}$, $Y:=KX$ linearizes every block except the quadratic term $\mu\overline\delta^2 X^2$ in $(1,1)$ and linear term $\mu I$ in $(5,5)$. Setting $\tau:=\mu^{-1}$ turn these into $(\overline\delta X)(\tau^{-1}I)(\overline\delta X)$ and $\tau^{-1}I$, respectively, and applying the Schur complement to the former appends a sixth block row and column, with $\overline\delta X$ at $(1,6)/(6,1)$ and $-\tau I$ at $(6,6)$, while $(5,5)$ remains $-\tau^{-1}I$. Finally, applying the congruence transformation $\diag(I,I,I,I,\tau I,I)$ maps $(4,5)/(5,4)$ from $I$ to $\tau I$ and $(5,5)$ from $-\tau^{-1}I$ to $-\tau I$.
\end{proof}

\subsection{Synchronization Result}

Consider the network of $N$ heterogeneous subsystems governed by the error dynamics \eqref{eq:error_dynamics}, where $e_i$ is the local tracking error, $w_i$ is a bounded external disturbance (as a consequence of Assumption \ref{ass:bounded_heterogeneity} and \ref{ass:leader_trajectory}), $d_i$ is the local topological coupling degree, and $\Delta_i$ represents parametric heterogeneity such that $\|\Delta_i\| \le \bar{\delta}$. Let $e = \operatorname{col}(e_1, \dots, e_N)$ and $w = \operatorname{col}(w_1, \dots, w_N)$ denote the global error and disturbance vectors.

The following theorem establishes conditions for practical synchronization under this unified dynamics and provides explicit ISS bounds.

\begin{theorem}[Practical Synchronization]
\label{th:main_theorem}
Consider the interconnected network \eqref{eq:error_dynamics}. Suppose there exist a common Lyapunov matrix $P \succ 0$, a control gain $K$, and positive scalars $\alpha$, $\gamma$, and $\eta$ such that the LMI \eqref{eq:CT_LMI_final} for continuous-time, or \eqref{eq:DT_LMI_final} for discrete-time, is feasible. If the network interconnection satisfies the small-gain condition:
\begin{equation}
    \label{eq:small_gain_condition}
    \gamma < \frac{\alpha}{\rho(D\mathcal{A})},
\end{equation}
where $D = \operatorname{diag}(d_1, \dots, d_N)$ and $\mathcal{A} = [a_{ij}]$, then the network achieves practical synchronization. The global error $e$ is Input-to-State Stable (ISS) with respect to $w$, satisfying:
\begin{equation}
    \|e(\mathfrak{t})\|^2 \le \kappa \beta(\mathfrak{t}) \|e(0)\|^2 + \zeta \sup_{s \in \mathcal{I}_\mathfrak{t}} \|w(s)\|_{\infty}^2,
\end{equation}
where $\|w\|_{\infty}=\max_i\|w\|$, $\mathfrak{t}$, $\beta(\mathfrak{t})$, and $\mathcal{I}_\mathfrak{t}$ are defined as:
\begin{itemize}
    \item Continuous-Time: $\mathfrak{t} = t \in\mathbb{R}_{\ge 0}$, $\beta(t) = e^{-\theta \alpha t}$, $\mathcal{I}_t = [0, t]$.
    \item Discrete-Time: $\mathfrak{t} = k \in\mathbb{Z}_{\ge 0}$, $\beta(k) = (1 - \theta \alpha)^k$, $\mathcal{I}_k = \{0, 1, \dots, k-1\}$.
\end{itemize}
Here $c \in \mathbb{R}^N$ is any vector such that $\frac{\gamma}{\alpha} D\mathcal{A}\, c < c$; and
\begin{align*}
    \theta &= \min_i \left(1 - \frac{\gamma}{\alpha}\frac{d_i \sum_{j\neq i} a_{ij}c_j}{c_i}\right) \in (0,1], \\
    \kappa &= \frac{N c_{\max} \lambda_{\max}(P)}{c_{\min} \lambda_{\min}(P)}, \\
    \zeta &= \frac{N c_{\max} \eta}{\theta \alpha c_{\min} \lambda_{\min}(P)},
\end{align*}
    
with $c_{\max} = \max_i c_i$, $c_{\min} = \min_i c_i$.
\end{theorem}

\begin{proof}
Let $V_i(e_i) = e_i^\top P e_i$ be the local ISS-Lyapunov candidate function. The feasibility of \eqref{eq:CT_LMI_final}--\eqref{eq:DT_LMI_final} guarantees the local dissipation inequality:
\[
    \delta V_i \le -\alpha V_i + \gamma d_i \sum_{j \neq i}^N a_{ij} V_j + \eta \|w_i\|^2.
\]
The condition $\gamma < \frac{\alpha}{\rho(D\mathcal{A})}$ implies 
$\rho\!\left(\frac{\gamma}{\alpha}D\mathcal{A}\right)<1$. Since 
$\frac{\gamma}{\alpha}D\mathcal{A}\ge0$ entrywise, this guarantees 
(see Remark \ref{rem:construction_c}) the existence of a strictly positive 
vector $c=[c_1,\dots,c_N]^\top>0$ such that
\[
    \frac{\gamma}{\alpha}D\mathcal{A}c<c .
\]
Equivalently, the above inequality holds componentwise as
\[
    \frac{\gamma}{\alpha}d_i\sum_{j\neq i}a_{ij}c_j<c_i,
    \qquad i=1,\dots,N .
\]
Therefore, we may define
\[
    \theta:=\min_i\left(1-\frac{\gamma}{\alpha}
    \frac{d_i\sum_{j\neq i}a_{ij}c_j}{c_i}\right)\in(0,1],
\]
which yields the bound
\begin{equation}
    \gamma d_i \sum_{j\neq i}^N a_{ij} c_j
    \;\le\; \alpha(1-\theta)c_i,
    \qquad i=1,\dots,N .
    \label{eq:interconnection_bound}
\end{equation}

We construct the global ISS-Lyapunov function $V(e) = \max_{i} \{V_i(e_i)/c_i\}$. Given the bounds $\lambda_{\min}(P)\|e_i\|^2 \le V_i \le \lambda_{\max}(P)\|e_i\|^2$, the global function satisfies:
\[
    \frac{\lambda_{\min}(P)}{N c_{\max}} \|e\|^2 \le V(e) \le \frac{\lambda_{\max}(P)}{c_{\min}} \|e\|^2.
\]
Let $m$ be the index of the active subsystem such that $V(e) = V_m(e_m)/c_m$. The unified variation $\delta V(e)$ satisfies:
\begin{align*}
    \delta V(e) &= \frac{\delta V_m}{c_m} \le -\alpha \frac{V_m}{c_m} + \frac{\gamma d_m}{c_m} \sum_{j \neq m}^N a_{mj} V_j + \frac{\eta}{c_m} \|w_m\|^2 \\
    &\le -\alpha V(e) + \gamma d_m \sum_{j \neq m}^N a_{mj} \frac{c_j}{c_m} V(e) + \frac{\eta}{c_{\min}} \|w\|_{\infty}^2 \\
    &\le -\alpha V(e) + \alpha(1-\theta) V(e) + \frac{\eta}{c_{\min}} \|w\|_{\infty}^2 \\
    &= -\theta \alpha V(e) + \frac{\eta}{c_{\min}} \|w\|_{\infty}^2,
\end{align*}
where the second to last step uses $V_j \le c_j V(e)$ for all $j$ (by definition of $V$ as the max) together with \eqref{eq:interconnection_bound} divided by $c_m$.

In the continuous-time case, $V$ is locally Lipschitz but non-differentiable where the active index $m$ switches; since the bound above holds for every active $V_i$, it holds for every element of the Clarke generalized gradient of $V$, exactly as argued in the proof of \cite[Thm. 11]{dashkovskiy2007iss}, to which we refer for details.

\textit{Continuous-Time Case ($\delta = d/dt$):} The inequality $\dot{V} \le -\theta \alpha V + \frac{\eta}{c_{\min}} \|w\|_{\infty}^2$ is a linear differential inequality. By the Comparison Lemma:
\begin{multline*}
    V(e(t)) \le e^{-\theta \alpha t} V(e(0)) + \frac{\eta}{c_{\min}} \int_0^t e^{-\theta \alpha (t-s)} \|w(s)\|^2 ds \le\\ e^{-\theta \alpha t} V(e(0)) + \frac{\eta}{\theta \alpha c_{\min}} \sup_{0 \le s \le t} \|w(s)\|^2.
\end{multline*}

\textit{Discrete-Time Case ($\delta V = V(k+1)-V(k)$):} The inequality $V(k+1) \le (1-\theta \alpha)V(k) + \frac{\eta}{c_{\min}} \|w\|_{\infty}^2$ implies:
\[
    V(k) \le (1-\theta \alpha)^k V(0) + \frac{\eta}{c_{\min}} \sum_{s=0}^{k-1} (1-\theta \alpha)^{k-1-s} \|w(s)\|^2.
\]
Bounding the geometric series $\sum_{j=0}^{k-1} (1-\theta \alpha)^j \le \frac{1}{\theta \alpha}$, we obtain:
\[
    V(k) \le (1-\theta \alpha)^k V(0) + \frac{\eta}{\theta \alpha c_{\min}} \sup_{0 \le s \le k-1} \|w(s)\|^2.
\]
Substituting the global norm bounds for $V(e)$ into the respective time-domain solutions yields the final ISS estimates for both domains.
\end{proof}

\begin{remark}[Optimal Selection of $c$]
\label{rem:optimal_c}
Theorem \ref{th:main_theorem} holds for any $c>0$ satisfying $\frac{\gamma}{\alpha}D\mathcal{A}c<c$, but $\theta$ depends on this choice and governs the convergence rate. Writing $\Gamma:=\frac{\gamma}{\alpha}D\mathcal{A}\ge 0$, we have $\theta(c) = 1-\max_i (\Gamma c)_i/c_i$, so maximizing $\theta$ is equivalent to the minimization $\rho(\Gamma) = \min_{c>0}\max_i (\Gamma c)_i/c_i$.

\textit{Irreducible case.} If $\mathcal{A}$ induces a strongly connected graph, $\Gamma$ is irreducible and admits a strictly positive Perron eigenvector $v>0$ with $\Gamma v=\rho(\Gamma)v$. Choosing $c=v$ yields the optimal rate $\theta^\star=1-\rho(\Gamma)=1-\frac{\gamma}{\alpha}\rho(D\mathcal{A})$, matching the residual margin in the small-gain condition.

\textit{Reducible case.} If $\mathcal{A}$ is reducible, the Perron eigenvector of $\Gamma$ may have zero entries and is thus inadmissible. We instead perturb: for $\varepsilon>0$, let $\Gamma_\varepsilon:=\Gamma+\varepsilon\mathbf{1}\mathbf{1}^\top$, which is irreducible and admits $v_\varepsilon>0$ with $\Gamma_\varepsilon v_\varepsilon=\rho(\Gamma_\varepsilon)v_\varepsilon$, where $\rho(\Gamma_\varepsilon)\to\rho(\Gamma)$ as $\varepsilon\to0^+$. Since $\Gamma\le \Gamma_\varepsilon$ entrywise, $\Gamma v_\varepsilon\le\rho(\Gamma_\varepsilon)v_\varepsilon$, so $c=v_\varepsilon$ is admissible and gives $\theta(v_\varepsilon)\ge 1-\rho(\Gamma_\varepsilon)$, arbitrarily close to $\theta^\star$ for small $\varepsilon$.
\end{remark}

The proposed synthesis requires only high-level knowledge of the network topology through the quantities entering the small gain condition ($\rho(D\mathcal{A})$) and the LMIs ($d_i$). If exact network information is unavailable, conservative estimates of the graph parameters can be employed, at the price of increased conservatism in the achievable performance bounds. On the other hand, this conservatism provides additional robustness against uncertainties in the network topology, ensuring that the synchronization guarantees remain valid even when the actual interconnection differs from the nominal one. It is worth emphasizing that the Perron-Frobenius vector is only required for the stability analysis and does not enter the controller implementation.

The ISS gain $\zeta$ explicitly quantifies the steady-state synchronization accuracy achieved by the network. In particular, since $\theta\ge 1-\frac{\gamma}{\alpha}\rho(D\mathcal A)$, the gain $\zeta$ depends both on design parameters ($P$, $\alpha$, $\gamma$, $\eta$) and on the network topology ($\mathcal{A}$). A larger decay rate~$\alpha$, a smaller disturbance gain~$\eta$, and a larger stability margin~$\theta$ all reduce $\zeta$, leading to tighter synchronization. 
The gain $\zeta$ can be made arbitrarily small only for CT dynamics, as we discuss in the following remark.

\begin{remark}[Continuous vs. Discrete-Time Asymmetry and Structural Limitations]
For a generic LTI system $\delta e = A_{\text{cl}} e + w$, subject to a persistent disturbance $w$, steady-state attenuation is governed by the static sensitivity matrix: $-A_{\text{cl}}^{-1}$ in continuous-time (CT) and $(I - A_{\text{cl}})^{-1}$ in discrete-time (DT). In CT, shifting eigenvalues arbitrarily far into the left half-plane ($\Re(\lambda) \to -\infty$) drives $\|A_{\text{cl}}^{-1}\| \to 0$, enabling arbitrary disturbance attenuation. Conversely, in DT, asymptotic stability strictly confines eigenvalues inside the unit circle ($|\lambda| < 1$), bounding the spectral distance $|1 - \lambda| < 2$. Thus, $\|(I - A_{\text{cl}})^{-1}\|$ cannot be made arbitrarily small, imposing an unavoidable structural floor on DT accuracy.

This principle applies directly to our collective error dynamics (25), governed by the open-loop pair $(I_N \otimes A) + \Delta = \operatorname{blkdiag}(A_1, \dots, A_N)$ and $L_{22} \otimes B$. While arbitrary error reduction via an unrestricted controller only requires standard controllability of this pair, our protocol enforces a replicated, block-diagonal structure $I_N \otimes K$.
Because a single static gain $K$ must simultaneously stabilize and tune the decay rate $\alpha$ across all heterogeneous agents, driving the CT error bound $\zeta$ in Theorem~1 to zero is far more restrictive than standard controllability, which is the price to pay for obtaining a synthesis independent of the network size. It requires either full actuation ($\operatorname{rank}(B) = n$) or matched heterogeneity $A_i = A_0 + B \tilde\Delta_i$ for some matrices $\tilde\Delta_i \in \mathbb{R}^{m \times n}$. Even when these stringent structural conditions are met, the DT synchronization error remains inherently bounded due to the unit-circle confinement.
\end{remark}

Theorem~\ref{th:main_theorem} provides an ISS bound with respect to the aggregate disturbance $w$. Since, in the synchronization problem, this disturbance is induced by the mismatch between the follower and leader dynamics, the result can be specialized directly to the leader state as follows.

\begin{corollary}
    Under the condition of Theorem~\ref{th:main_theorem}, the aggregate error system \eqref{eq:global_error_system} is ISS with respect to the leader state $x_0$, in the sense of Definition~\ref{def:practical_sync}. In particular,
    \begin{equation}
        \label{eq:iss_x0_bound}
        \|e(\mathfrak{t})\|^2 \le \kappa \beta(\mathfrak{t})\|e(0)\|^2 + 4\bar\delta^2\zeta \sup_{s\in\mathcal{I}_\mathfrak{t}}\|x_0(s)\|^2
    \end{equation}
    with $\mathfrak{t}$, $\beta(\mathfrak{t})$, $\mathcal{I}_\mathfrak{t}$ as in Theorem~\ref{th:main_theorem}.
\end{corollary}
\begin{proof}
    By construction $w_i=H_i x_0$ with $H_i=\Delta_i-\Delta_0$. Under Assumption~\ref{ass:bounded_heterogeneity}, $\|H_i\|\le\|\Delta_i\|+\|\Delta_0\|\le 2\bar\delta$ for all $i$, so $\|w_i(s)\|\le 2\bar\delta\|x_0(s)\|$ for all $s\in\mathbb{T}$, hence $\|w(s)\|_\infty \le 2\bar\delta\|x_0(s)\|$. Substituting into the bound of Theorem~\ref{th:main_theorem} gives \eqref{eq:iss_x0_bound}.
\end{proof}

This bound makes explicit how the leader trajectory and the heterogeneity level determine the ultimate synchronization accuracy.

\section{Numerical Example}
\label{section:numerical_example}

We validate the theoretical results of Section~\ref{sec:main_results} on a synchronization problem for a heterogeneous network of linear harmonic oscillators, comparing open-loop and closed-loop trajectories and the average synchronization error for both the Continuous-Time (CT) and Discrete-Time (DT) designs. We further show that a discretized CT controller loses stability under sampling, motivating the native DT synthesis of Proposition~\ref{prop:DT_LMI}.

\subsection{Network Description}

\begin{figure}[htpb]
    \centering
    \resizebox{0.8\columnwidth}{!}{%
    \begin{tikzpicture}[
        every node/.style={
            circle,
            draw,
            align=center,
            minimum size=1.2cm,
            font=\small
        },
        >=Stealth,
        scale=0.9
    ]

    \node[fill={rgb,1:red,0.85;green,0.35;blue,0.35}, thick] (1) at (0,5) {\textbf{(0)} \\ 50 Hz};

    \node[fill={rgb,1:red,0.85;green,0.65;blue,0.45}] (2) at (-4,3) {\textbf{(1)} \\ 53 Hz};
    \node[fill={rgb,1:red,0.90;green,0.82;blue,0.55}] (3) at (4,3) {\textbf{(2)} \\ 41 Hz};

    \node[fill={rgb,1:red,0.70;green,0.50;blue,0.85}] (7) at (-2,1) {\textbf{(6)} \\ 56 Hz};
    \node[fill={rgb,1:red,0.55;green,0.80;blue,0.55}] (4) at (2,1) {\textbf{(3)} \\ 60 Hz};
    \node[fill={rgb,1:red,0.60;green,0.88;blue,0.95}] (5) at (-2,-1) {\textbf{(4)} \\ 40 Hz};
    \node[fill={rgb,1:red,0.85;green,0.45;blue,0.70}] (6) at (2,-1) {\textbf{(5)} \\ 42 Hz};

    \draw[->] (1) -- (2);
    \draw[->] (1) -- (3);
    \draw[->] (1) to[bend left=15] (4);

    \draw[->] (2) to[bend left=10] (4);
    \draw[->] (2) to[bend right=10] (5);

    \draw[->] (3) to[bend left=5] (6);
    \draw[->] (3) .. controls (4.5,-2.2) and (1.5,-3.2) .. (5);

    \draw[->, very thick, red, rounded corners=8pt] (5) -- (7);
    \draw[->, very thick, red, rounded corners=8pt] (7) -- (4);
    \draw[->, very thick, red, rounded corners=8pt] (4) -- (6);
    \draw[->, very thick, red, rounded corners=8pt] (6) -- (5);

    \end{tikzpicture}
    }
    \caption{Directed communication graph topology $\mathcal{G}$ representing the interaction between the virtual leader (Agent 0) and the $N=6$ heterogeneous followers.}
    \label{fig:network_topology}
\end{figure}
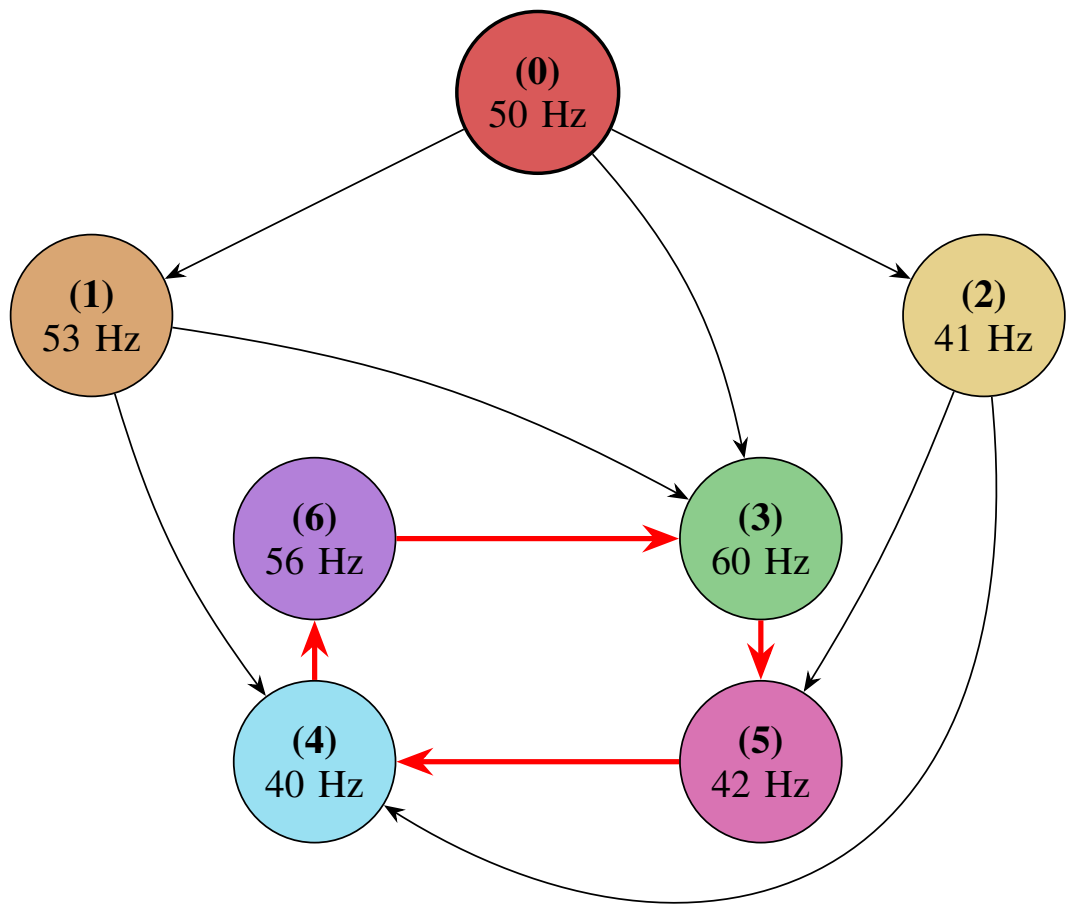

Consider $N=6$ follower agents and a virtual leader (Agent 0), modeled as heterogeneous harmonic oscillators. The leader oscillates at $f_0 = 50\,\text{Hz}$ ($\omega_0 = 100\pi\,\text{rad/s}$), while the followers' natural frequencies $f_i$ are drawn uniformly from $[40,60]\,\text{Hz}$; the nominal plant matrix $A$ is defined from the leader's dynamics, with heterogeneity bound $\bar\delta = \max_i \|\Delta_i\|$. The leader-follower interaction is modeled by the directed graph $\mathcal G$ in Fig.~\ref{fig:network_topology}, which contains a spanning tree rooted at the leader.

\subsection{Continuous-Time Design}

The CT local dynamics are
\begin{equation}
    \label{eq:oscillator_models}
    \dot{x}_i = \begin{bmatrix}
        0 & \omega_i\\ -\omega_i & 0
    \end{bmatrix}x_i + \begin{bmatrix}
        0\\ 1
    \end{bmatrix}u_i.
\end{equation}
Applying Proposition~\ref{prop:CT_LMI} with $\bar\delta = 62.8319$ yields the stabilizing gain $K_{\text{CT}} = \begin{bmatrix}-479.5712 & -906.0123\end{bmatrix}$, with $P = 10^7\begin{bmatrix}7.6550 & 3.9259\\ 3.9259 & 7.3685\end{bmatrix}$, $\gamma = 2.0368$, $\eta = 6.5076\times10^4$, $\alpha = 11.0009$, and $\tau = 7.2585\times10^{-7}$, satisfying the small-gain condition \eqref{eq:small_gain_condition} of Theorem~\ref{th:main_theorem} and thus guaranteeing practical synchronization.

\begin{figure}[htpb]
    \centering
        \includegraphics[width=\columnwidth]{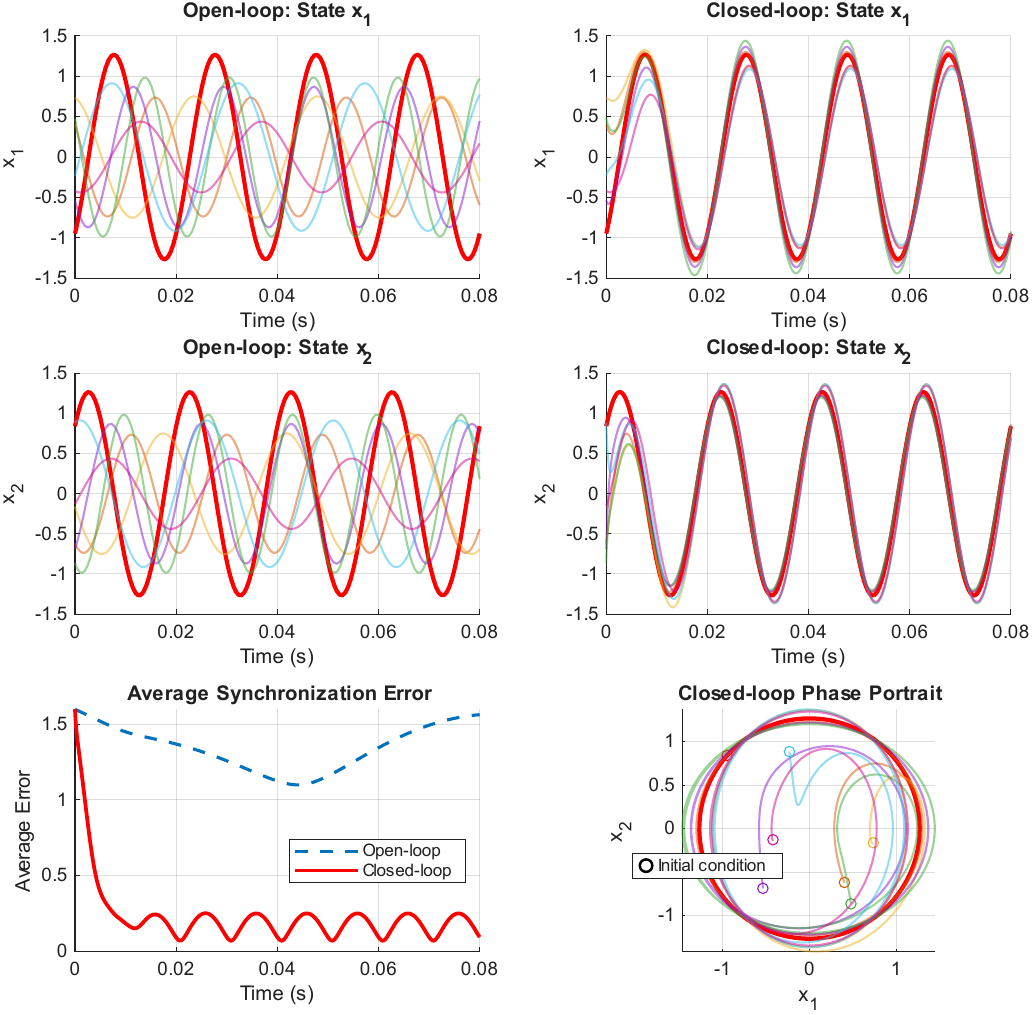}
    \caption{Comparison of open-loop and closed-loop state evolution $x_i(t)$ for the Continuous-Time (CT) design. Trajectory colors correspond to the node coloring defined in Fig.~\ref{fig:network_topology}, and the virtual leader (Agent 0) is highlighted in red. The average synchronization error $\bar e(t)$ is also shown for the open-loop and closed-loop CT designs.}
    \label{fig:ct_avg_error}
\end{figure}

Fig.~\ref{fig:ct_avg_error} compares the open-loop network (each follower evolving at its own mismatched frequency, $u_i\equiv0$) with the closed-loop network under $K_{\text{CT}}$: the open-loop agents drift apart, while the closed-loop agents lock to the leader's oscillation. The corresponding average error $\bar e(t) := \frac1N\sum_i\|e_i(t)\|$ confirms this, converging to a bounded neighborhood of the origin in closed loop, consistent with Theorem~\ref{th:main_theorem}.

\subsection{Discrete-Time Design}
\label{subsec:dt_design}

For a sampling frequency $f_s = 1\,\text{kHz}$ ($T_s = 1\,\text{ms}$), we use the native DT framework of Proposition~\ref{prop:DT_LMI}. Exact ZOH discretization of \eqref{eq:oscillator_models} gives $A_{d,i} = e^{A_iT_s}$, $B_{d,i} = \big(\int_0^{T_s}e^{A_i\tau}d\tau\big)B$; since $A_i$ depends on $\omega_i$, the exact $B_{d,i}$ are agent-dependent, incompatible with the common-input-matrix structure of Section~\ref{sec:main_results}. We therefore evaluate $B_d$ at the leader's frequency $\omega_0$ and absorb the mismatch into a scalar $\phi_i$ with $B_{d,i}\simeq\phi_i^{-1}B_d$, giving the synthesis model
\begin{equation*}  \label{eq:dt_oscillator_models}
    \resizebox{\columnwidth}{!}{$
    x_i[k+1] = \begin{bmatrix}
        \cos(\omega_i T_s) & \sin(\omega_i T_s) \\
        -\sin(\omega_i T_s) & \cos(\omega_i T_s)
    \end{bmatrix} x_i[k] + \begin{bmatrix}
        \frac{1-\cos(\omega_0 T_s)}{\omega_0} \\
    \frac{\sin(\omega_0 T_s)}{\omega_0}
    \end{bmatrix} u_i[k],
    $}
\end{equation*}
with implemented control law $u_i = \phi_i K_{\text{DT}} \sum_{j} a_{ij}(x_i - x_j)$, i.e., individualized gains $K_i = \phi_i K_{\text{DT}}$ compensating for the input mismatch while retaining the common Lyapunov certificate $P$ synthesized in Proposition~\ref{prop:DT_LMI}.

Applying Proposition~\ref{prop:DT_LMI} with $\bar\delta=0.0628$ yields $K_{\text{DT}} = -\begin{bmatrix}11.5933 & 221.8548\end{bmatrix}$, with $P = 10^4\begin{bmatrix}4.9520 & 0.9400\\ 0.9400 & 4.0458\end{bmatrix}$, $\gamma=0.0041$, $\eta=5.5644\times10^4$, $\alpha=0.1100$, $\tau=6.5356\times10^{-6}$, again satisfying \eqref{eq:small_gain_condition} and guaranteeing practical synchronization.

\begin{figure}[htpb]
    \centering
    \includegraphics[width=\columnwidth]{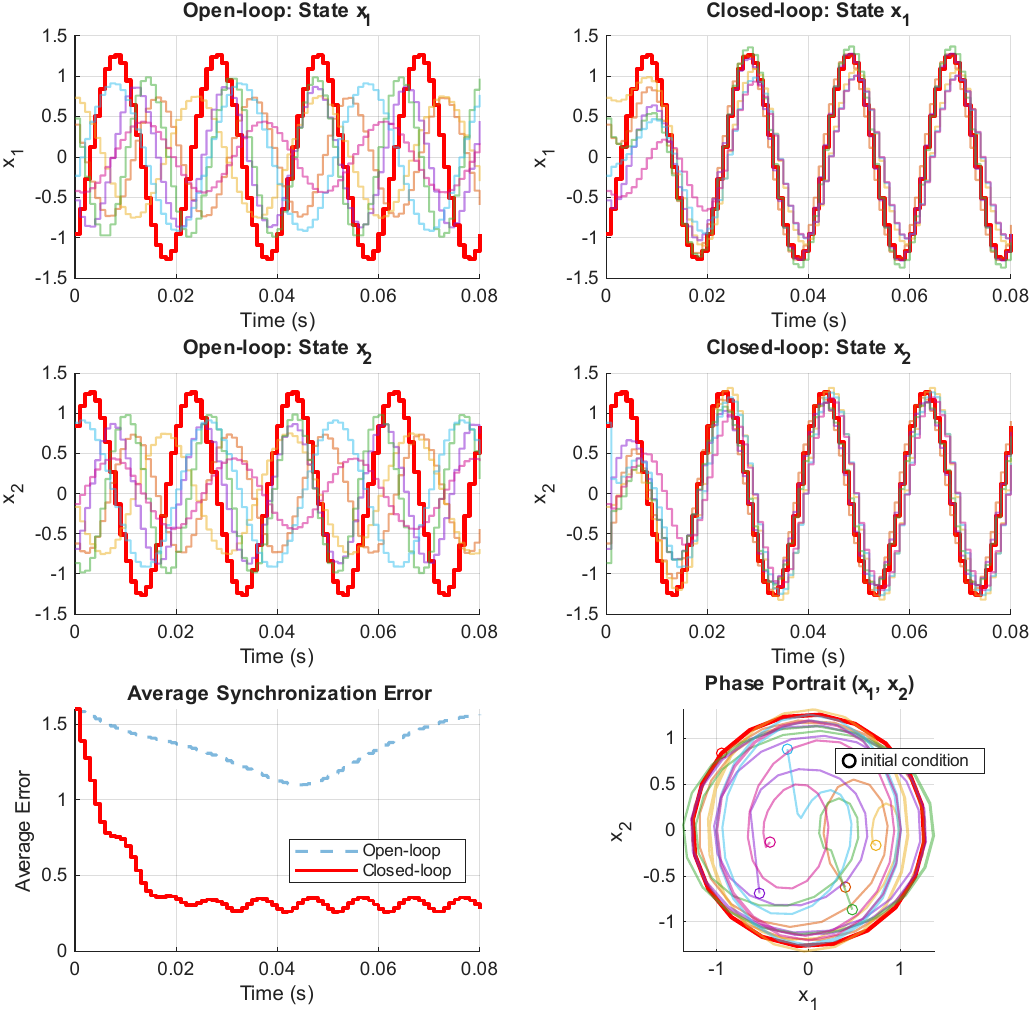}
    \caption{Comparison of open-loop and closed-loop state evolution $x_i[k]$ for the native Discrete-Time (DT) design at $f_s = 1\,\text{kHz}$. Trajectory colors correspond to the node coloring defined in Fig.~\ref{fig:network_topology}, and the virtual leader (Agent 0) is highlighted in red. The average synchronization error $\bar e[k]$ is also shown for the open-loop and closed-loop DT designs.}
    \label{fig:dt_avg_error}
\end{figure}

As in the CT case, Fig.~\ref{fig:dt_avg_error} compares open-loop ($u_i[k]\equiv0$) and closed-loop trajectories $x_i[k]$, along with the average error $\bar e[k] := \frac1N\sum_i\|e_i[k]\|$: the native DT design drives $\bar e[k]$ to a bounded neighborhood of the origin, whereas in open loop the followers drift apart at their mismatched frequencies.

\subsection{Discrete Implementation and CT Design Limitations}
\label{subsec:ct_limitations}

Implementing the CT synthesis via ZOH discretization alters the plant's spectral properties. 

\begin{figure}[htpb]
    \centering
    \begin{subfigure}[b]{0.98\linewidth}
        \centering
        \includegraphics[width=\linewidth]{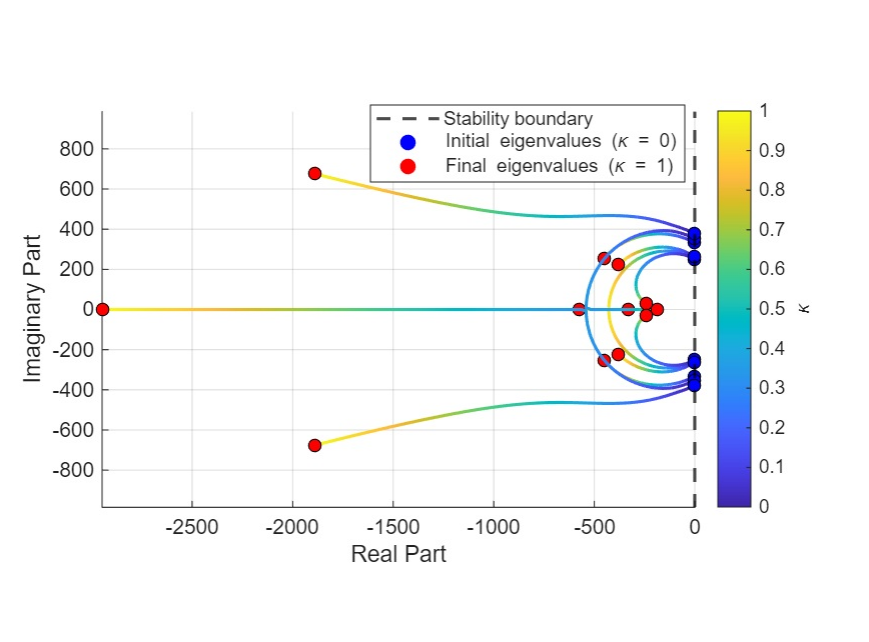}
        \caption{CT Plant ($A_i, B$) with CT Gain $K_{\text{CT}}$}
        \label{fig:eig_ct}
    \end{subfigure}
    \hfill
    \begin{subfigure}[b]{0.98\linewidth}
        \centering
        \includegraphics[width=\linewidth]{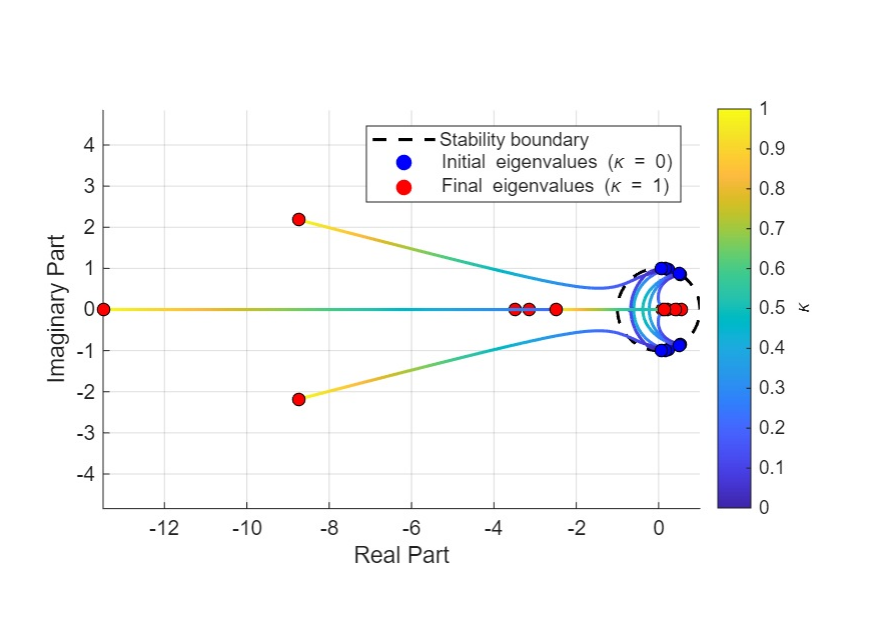}
        \caption{DT Plant ($A_{d,i}, B_{d,i}$) with CT Gain $K_{\text{CT}}$}
        \label{fig:eig_dt_kct}
    \end{subfigure}
    \hfill
    \begin{subfigure}[b]{0.98\linewidth}
        \centering
        \includegraphics[width=\linewidth]{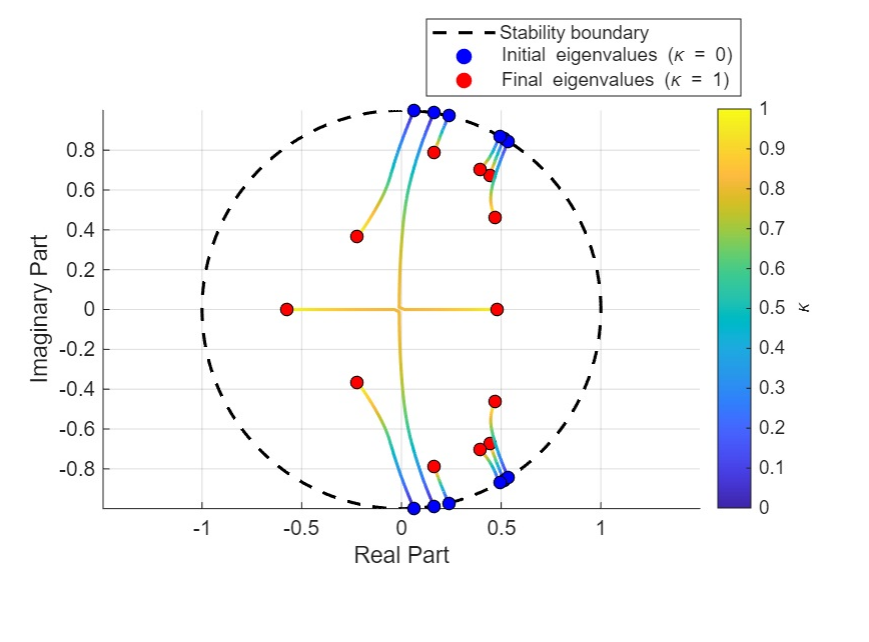}
        \caption{DT Plant ($A_{d,i}, B_{d,i}$) with Scaled DT Gain $\phi_i K_{\text{DT}}$}
        \label{fig:eig_dt_kdt}
    \end{subfigure}
    \caption{Comparison of the closed-loop eigenvalue distribution for the synchronization error dynamics as $\kappa \in [0, 1]$ varies. The native DT design (c) inherently accounts for sample-and-hold dynamics and maintains stability inside the unit circle, whereas the discretized CT design (b) fails to preserve stability.}
    \label{fig:eigenvalues_comparison}
\end{figure}

To illustrate this, we analyze the closed-loop spectrum of the aggregate error dynamics by varying the gain scaling factor $\kappa \in [0, 1]$ in the expression $(I_{N} \otimes A) + \Delta + (L_{22} \otimes \kappa BK)$. As depicted in Fig.~\ref{fig:eigenvalues_comparison}, we compare the eigenvalue distribution under three distinct scenarios:

\begin{itemize}
    \item Ideal CT Design (Fig.~\ref{fig:eigenvalues_comparison}a): The spectrum of the continuous closed-loop matrix using the exact continuous plant and the direct CT gain $K_{\text{CT}}$. The eigenvalues for $\kappa = 1$ lie strictly in the left half-plane, confirming the stability of the CT synthesis.
    
    \item Discretized CT Design (Fig.~\ref{fig:eigenvalues_comparison}b): The spectrum of the true discretized plant $(A_{d,i}, B_{d,i})$ evaluated with the discretized CT controller $K_{\text{CT}}$. At $\kappa = 1$, several eigenvalues migrate outside the unit circle, indicating that the CT controller fails to stabilize the discrete plant.
    
    \item Native DT Design (Fig.~\ref{fig:eigenvalues_comparison}c): The spectrum utilizing the exact discretized plant $(A_{d,i}, B_{d,i})$, same sampling frequency as in the previous example, and the normalized native DT controller $K_i = \phi_i K_{\text{DT}}$ derived from Proposition~\ref{prop:DT_LMI}. The eigenvalues for $\kappa = 1$ remain safely constrained within the unit circle, validating the robustness of the discrete-time synthesis that natively accounts for the forward-shift operator dynamics.
\end{itemize}

\section{Conclusions}

\label{section:conclusion}

    We developed a network-size-independent LMI synthesis procedure for practical leader-following synchronization of heterogeneous linear systems under static diffusive state feedback. By treating the mismatch-induced terms as disturbances, local robust dissipation inequalities were combined through an ISS small-gain construction to obtain explicit global synchronization-error bounds. The unified formulation also reveals a qualitative CT/DT asymmetry: while increasing closed-loop bandwidth can reduce the effect of persistent disturbances in continuous time under suitable actuation and matching conditions, discrete-time stability imposes a finite limit on generic disturbance attenuation. Numerical experiments further show that a controller designed in continuous time need not remain stabilizing after ZOH discretization, supporting the need for native discrete-time synthesis.

    The price paid for scalability is conservatism, arising primarily from the common Lyapunov matrix and the robust treatment of the coupling degree and plant mismatch. An important direction for future work is therefore to reduce this conservatism while retaining fixed-dimension synthesis, for example through parameter-dependent Lyapunov functions or structured treatment of heterogeneous input matrices.  Extensions to nonlinear systems and to time-varying topologies are also of interest.



\bibliographystyle{abbrv}
\bibliography{bibliography}

\begin{IEEEbiography}%
[{\includegraphics[width=1in,height=1.25in,clip,keepaspectratio]{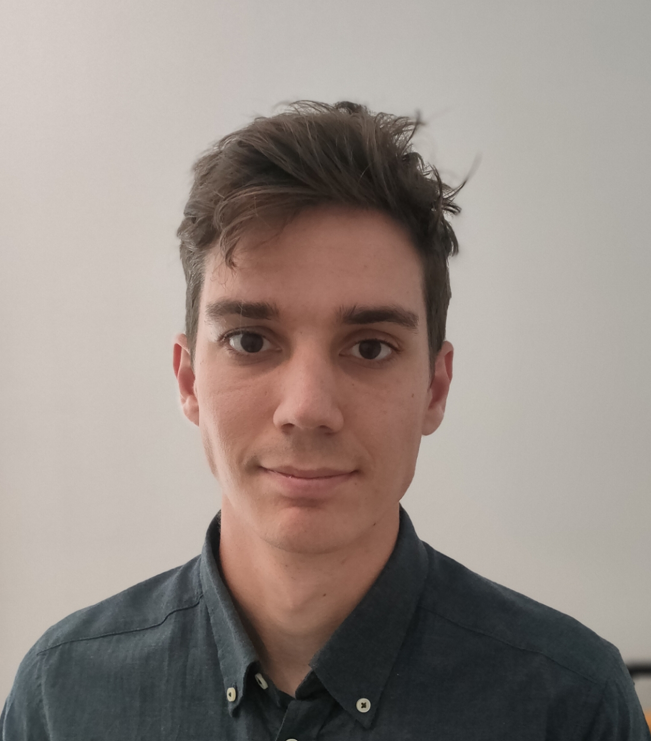}}]%
{Alex Ardelean}  received the Master degree in robotics engineering from
University of Perugia, Perugia, in 2025. 
He is currently a Ph.D.\ student in GIPSA-Lab, Grenoble, France. His research interests include multi-agent systems, specializing in synchronization of discrete time systems. 
\end{IEEEbiography}

\begin{IEEEbiography}
[{\includegraphics[width=1in,height=1.25in,clip,keepaspectratio]{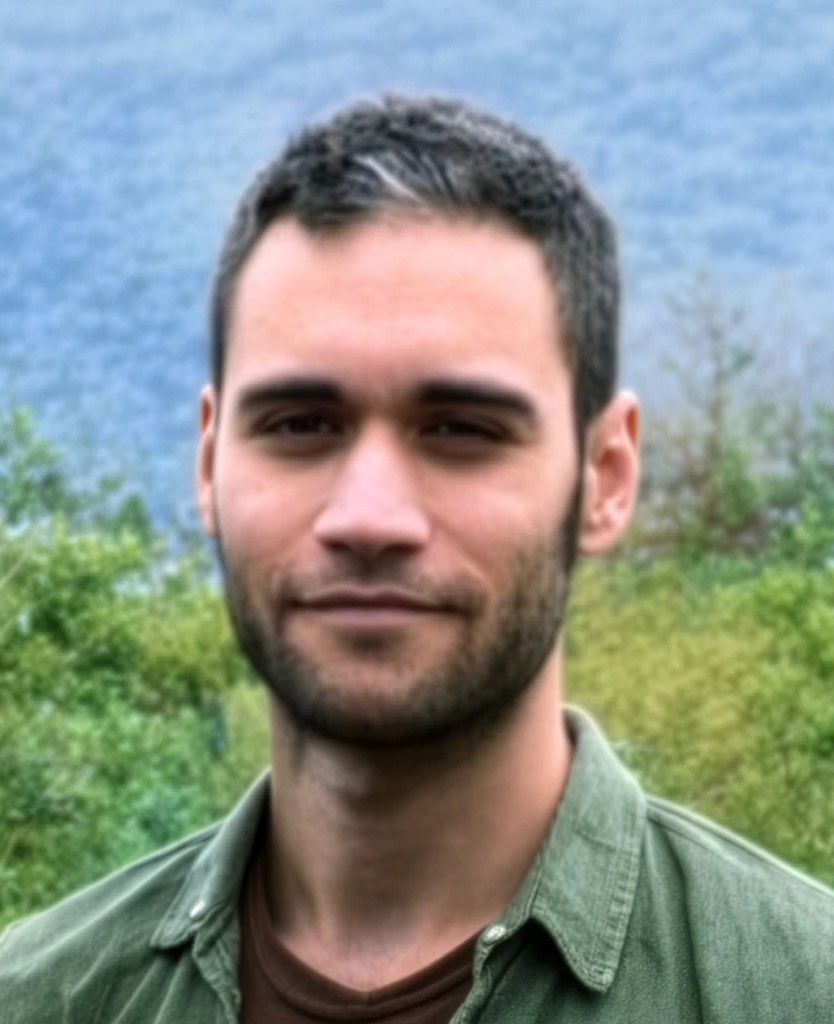}}]
{Giacomo Casadei} is an associate professor in Universitè Grenoble Alpes (IUT) and a member of team DANCE in GIPSA-Lab since 2024. He received the Master degree in Automation in 2012 and the Phd degree in Automation and Operative Research in 2016 at the University of Bologna, Italy. From 2018 to 2024, he has held an associate professor position in Ecole Centrale de Lyon. His research interests include modeling and control of networks, with a particular focus on synchronization of nonlinear systems and the use of nonlinear control techniques.
\end{IEEEbiography}

\begin{IEEEbiography}[{\includegraphics[width=1in,height =1.25in,clip,keepaspectratio]{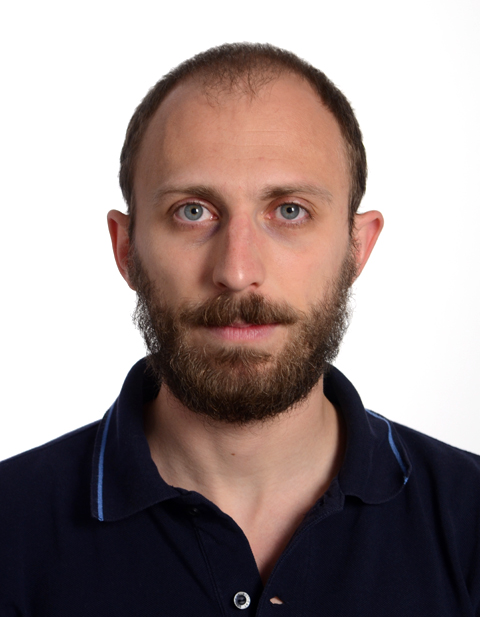}}]{Paolo Frasca} (Senior Member, IEEE) received the Ph.D. degree from the Politecnico di Torino, Turin, Italy, in 2009. 
He was an Assistant Professor at the University of Twente, Enschede, Netherlands, from 2013 to 2016. Since October 2016, he has been a CNRS Researcher at GIPSA-Lab, Grenoble, France, where since 2021 he has been leading the Dynamics and Control of Networks (DANCE) group. His research interests include control systems and network theory, with applications in transportation and social networks. Dr.\ Frasca has served as an Associate Editor for several conferences and journals, including the IEEE Control Systems Letters and Automatica.
\end{IEEEbiography}

\end{document}